# Morphological Comparison of Blocks in Chaos Terrains on Pluto, Europa, and Mars

Helle L. Skjetne[1*], Kelsi N. Singer[1], Brian M. Hynek[2], Katie I. Knight[3], Paul M. Schenk[4], Cathy B. Olkin[1], Oliver L. White[5], Tanguy Bertrand[6], Kirby D. Runyon[7], William B. McKinnon[8], Jeffrey M. Moore[6], S. Alan Stern[1], Harold A. Weaver[7], Leslie A. Young[1], and Kim Ennico[7]

[1]Southwest Research Institute, 1050 Walnut St. Suite 300, Boulder, CO 80302
[2]Labratory for Atmospheric and Space Physics, University of Colorado at Boulder, Boulder, CO
[3]Carson-Newman University, Jefferson City, TN
[4]Lunar and Planetary Institute, Houston, TX
[5]The SETI Institute, Mountain View, CA
[6]NASA Ames Research Center, Space Science Division, Moffett Field, CA
[7]Johns Hopkins Applied Physics Laboratory, Laurel, MD
[8]Department of Earth and Planetary Sciences and McDonnell Center for the Space Sciences, Washington University in St. Louis, 1 Brookings Dr., St. Louis, MO 63130, USA

**Running Head:**
Chaos Terrains Across the Solar System

*Corresponding Author:
Helle L. Skjetne
1050 Walnut St. Suite #300
Boulder, CO 80302
Hskjetne@vols.utk.edu
Phone: (720) 933-9903

# MORPHOLOGICAL COMPARISON OF BLOCKS IN CHAOS TERRAINS ON PLUTO, EUROPA, AND MARS

S<small>KJETNE ET AL</small>., 2021

## ABSTRACT

Chaotic terrains are characterized by disruption of preexisting surfaces into irregularly arranged mountain blocks with a "chaotic" appearance. Several models for chaos formation have been proposed, but the formation and evolution of this enigmatic terrain type has not yet been fully constrained. We provide extensive mapping of the individual blocks that make up different chaos landscapes, and present a morphological comparison of chaotic terrains found on Pluto, Jupiter's Moon Europa, and Mars, using measurements of diameter, height, and axial ratio of chaotic mountain blocks. Additionally, we compare mountain blocks in chaotic terrain and fretted terrain on Mars. We find a positive linear relationship between the size and height of chaos blocks on Pluto and Mars, whereas blocks on Europa exhibit a flat trend as block height does not generally increase with increasing block size. Block heights on Pluto are used to estimate the block root depths if they were floating icebergs. Block heights on Europa are used to infer the total thickness of the icy layer from which the blocks formed. Finally, block heights on Mars are compared to potential layer thicknesses of near-surface material. We propose that the heights of chaotic mountain blocks on Pluto, Europa, and Mars can be used to infer information about crustal lithology and surface layer thickness.

## HIGHLIGHTS

- We compared chaotic mountain blocks on Pluto, Europa, and Mars.
- We measured the block diameters, heights, and axial ratios.
- Mountain block heights on Pluto and Mars increase with increasing block size.
- In contrast europan block heights are generally similar, regardless of size for a given chaos region.
- Block heights can be used to infer crustal lithology and surface layer thickness.

# 1. INTRODUCTION

Chaos terrain is formed by disruption of preexisting surfaces. This typically occurs through fracturing (that can be induced through a variety of mechanisms), which produces an appearance of irregularly shaped blocks with a "chaotic" appearance (Meresse et al., 2008; Collins and Nimmo, 2009; Zegers et al., 2010; and see **Tables 1** and **2**; **Fig. 1**). After fractures develop, the subsequent evolution of these blocks can follow several paths. If the blocks are completely destabilized and free from the surface below they may rotate and translate. Block flotation may occur in a liquid or solid with sufficient density contrast and sufficiently low viscosities at the given surface temperatures, and may reach an isostatic position (Collins and Nimmo, 2009; Schmidt et al., 2011). Alternatively, the blocks may remain in place and the fractures around them may be deepened over time by erosion (e.g., on Mars; Rodriguez et al., 2005; Andrews-Hanna and Phillips, 2007; Warner et al., 2011). The total degree of disruption also varies. In some cases, the entire pre-existing terrain is effectively resurfaced by demolition into small-scale blocks (at or below the resolution limits; e.g., the "matrix" material in europan chaos as seen in **Fig. 1**) and no large blocks remain. The geologic evolution required to shape these distinct terrains is not fully understood, although several chaos formation models have been proposed (see **Section 2** for discussion).

The goal of this work is to provide more detailed information about the morphology of mountain blocks in chaos terrain across several solar system bodies by mapping chaotic terrain blocks. These distinctive areas of broken terrains are most notably found on Jupiter's moon Europa, Mars, and Pluto. Although chaos terrains on these bodies share some common characteristics, there are also distinct morphological differences between them. We explore the similarities and differences between the morphology of blocks in chaotic terrain in selected regions (see **Section 2**) across these three bodies.

Physical characteristics of chaotic terrains can be used to infer information about the crustal lithology of each planet (e.g., Williams and Greeley, 1998; Schmidt et al., 2011; Singer et al., 2021). We report on aspects of block morphologies, and also present size (diameter) and height (the maximum elevation exposed above the mean basal elevation, discussed in more detail in **Section 3.3**) distributions as a way to compare between bodies and constrain formation theories where possible. In addition, we also provide a comparison between martian chaos blocks and blocks found in fretted terrain along the lowland dichotomy boundary on Mars. Fretted terrain is another common martian lowland terrain type that shows similarities to martian chaotic terrain (both characterized by large steep-walled mesas and plateaus separated by flat-floored valleys; Carr, 2001). Providing comparison to similar terrain types across the solar system could provide constraints on the formation of chaotic terrain and the geologic evolution of each body.

Regional context and background are provided in **Section 2**. Mapping methods are summarized in **Section 3**, and the results are given in **Section 4**. **Section 5** compares the results to those of previous studies and discusses the implications of these results for crustal lithology on each body. A summary of our main findings and outlook for future work are given in **Section 6**.

# 2. REGIONAL CONTEXT AND BACKGROUND

We mapped chaos terrain blocks on three bodies: Pluto, Europa, and Mars. In addition, we also included martian fretted terrain blocks. On each body, the chaotic terrain was subdivided into regions based on geographical location and geological context. Note that some feature names used in this paper are informal.

## 2.1 Pluto

The chaotic mountain ranges (or "montes") on Pluto occur mainly along the western margin of the high albedo, nitrogen-rich ice sheet called Sputnik Planitia (SP) from ~45° N to ~25° S and 150° E to 180° E (Stern et al., 2015; White et al. 2017) as an intermittent series of angular blocks, reaching up to ~35 km across and ~4 km high above the surrounding nitrogen-ice rich plains (Moore et al., 2016). These larger blocks are surrounded by a finer textured chaotic inter-block material that could be the disintegrative product of chaotic mountain blocks breaking into smaller fragments (White et al., 2017). We mapped and studied the six primary mountain ranges on Pluto (from north to south): Al-Idrisi Montes, Zheng He Montes, Baret Montes, Hillary Montes, Tenzing Montes, and Tabei Montes. For descriptions of each region, see **Table 1** (see **Fig. A1** in **Appendix A** for the locations of each mountain range on Pluto).

The mountain ranges on Pluto appear to be broken blocks of the primarily water-ice lithosphere around the edge of SP. The blocks are tall topographic features, indicating they are made of water ice ($H_2O$) and possibly partly methane ice ($CH_4$), rather than nitrogen ice ($N_2$), which cannot retain significant topography at Pluto's surface temperatures (Moore et al., 2016). Several mountain blocks in the Al-Idrisi range also display hints of one or more alternating dark and light layers, which resemble layering found in the walls of craters in the nearby terrain outside of SP (White et al., 2017; Singer et al., 2021). Spectroscopic measurements from New Horizons have also confirmed the presence of $H_2O$ and $CH_4$ in the mountain block regions (Grundy et al., 2016; Cook et al., 2019). Although methane is a volatile ice that sublimates and re-deposits, forming thin surface deposits at many locations on Pluto, it may also be a part of the observed subsurface layers.

Sputnik Planitia, covering just over 11% of Pluto's surface area (Moore et al., 2016), is a complex area that features a wide variety of geologic phenomenon, including evidence for glacial flow of the nitrogen-rich ice (White et al., 2017). The SP nitrogen ice sheet, and the mountains within, are located in a very large, ancient impact basin — Sputnik basin. It is possible the fracturing of the pre-existing crust occurred as part of structural modification of the lithosphere from the basin forming impact, and the mountains may even be part of a multi-ring basin structure as the distribution of these mountain ranges fits within the $\sqrt{2}$ spacing of lunar basin rings (McKinnon et al., 2017). Additionally, a long, ~N–S trending, extensional fault system runs from the northern polar region down along the western edge of the Sputnik basin (White et al., 2017, Schenk et al., 2018), proximal to many of the mountain ranges with chaotic blocks, and continues southward. This fault system may have contributed to breakup of the northernmost mountain range Al-Idrisi Montes and is also possibly related to other montes regions as well (Schenk et al., 2018; **Table 1**). There are a few instances of smaller fractures appearing near the rim of SP (see White et al., 2019; their Figs. 5 and 11), that may represent partially fractured or currently fracturing blocks (Howard et al., 2017; White et al., 2017).

The chaotic mountain ranges emerge from cellular plains that consists of volatile ice mixtures, primarily $N_2$, but also some carbon monoxide (CO) and $CH_4$ (Moore et al., 2016; Grundy et al., 2016; Protopapa et al. 2017; Schmitt et al., 2017). Pure water ice and nitrogen ice may have a density contrast of >5% at Pluto's average surface temperature of ~40 K (Scott, 1976). It is possible the chaotic blocks could have been partially or fully floating icebergs in the SP nitrogen ice sheet (discussed at length in the supplement online materials of Moore et al., 2016), which could assist with destabilization or fracturing/tilting. Bertrand et al. (2018) estimated that a net amount of 1 km ice has sublimed at the northern end of SP during the last millions of years, and this removal could contribute to driving viscous flow of the ice sheet. Glacial flow of $N_2$ ice off the surrounding highlands or within SP itself may have assisted with breakup and translation of the blocks even if they were not completely buoyant (Howard et al., 2017; White et al., 2017).

**Table 1. Descriptive summary of regions studied on Pluto.** Base mosaic resolution: ~315 m px$^{-1}$ (see Section 3.1 for further information).

| Region | Description | Image |
|---|---|---|
| *Al-Idrisi Montes* (~30° N, 154° E) | This region is located in the upper NW margin of SP between the bright cellular plains of SP and the uplands surrounding SP. The region features large blocks surrounded by chaotic inter-block material. Some of the blocks appear to have been tilted and rotated (assuming that the surfaces with textures resembling the upland terrain represent their original tops), and the blocks do not appear to have a preferred dip orientation (see Appendix C for slope characterization of tilted blocks in regions for Pluto). In the northern extent of the region the blocks generally become smaller in size, and merge into chaotic inter-block material in the region where EW-trending extensional faults are located (White et al., 2017). | 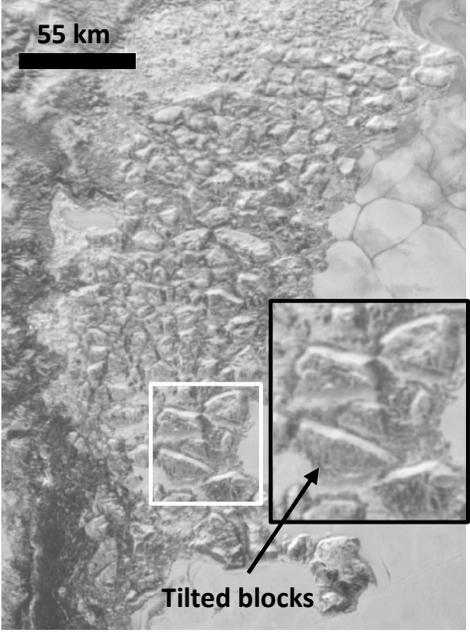 |
| *Baret Montes* and *Zheng He Montes* (~15° N, 158° E) | This region features blocks in Baret Montes and Zheng He Montes. Although blocks in the Baret Montes region appear to have only been minimally tilted compared to blocks in the Zheng He Montes region, most blocks appear to feature roughly "flat-topped" surfaces (including presumably tilted blocks in Zheng He Montes) that resemble the texture of the terrain found to the west outside of SP. The lower section of Baret Montes is connected to a dark ridged terrain with a curving compression pattern trending N–S or NW–SW (White et al., 2017). It is possible that this dark ridged terrain could be less fractured terrain of the same composition as the mountain blocks found in Baret Montes. | 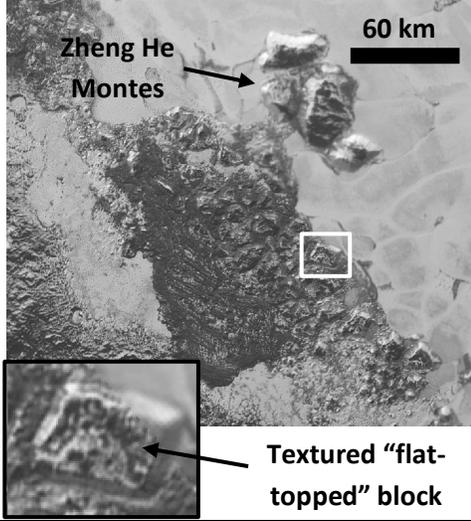 |
| *Hillary Montes* (~5° N, 168° E) | This region features blocks with a similar morphology to blocks in the Baret Montes region, although there is generally more N$_2$ ice or smaller, inter-block matrix-like material separating them. The tops of the blocks (where they do not appear angular) in this region resemble the texture of the terrain located to the west of the region. The tops of blocks typically appear more angular (forming distinct peaks) in the southern extent of the region. | 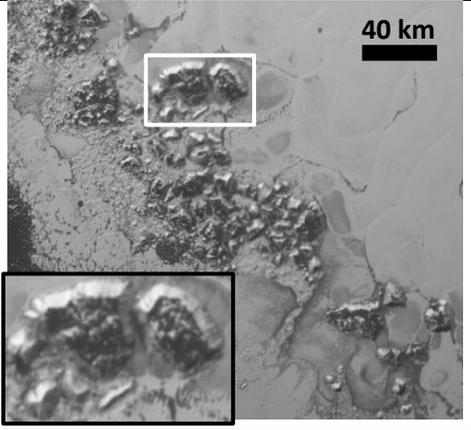 |
| *Tenzing Montes* (~15° S, 177° E) | This region features blocks with a significantly more angular and "spikey" appearance compared to blocks in other regions. Some blocks display a highly tilted appearance. The texture of the apparent block "tops" resembles the texture of the surrounding terrain found to the west of the region outside SP. | 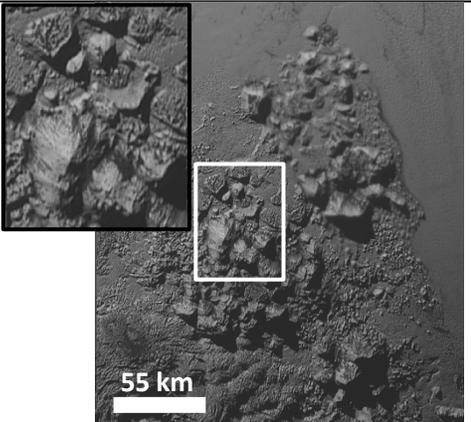 |
| *Tabei Montes* (~10° S, 165° E) | This region consists of ~2 areas of tightly clustered blocks with a sub-rounded appearance. This region is unique because it is the only collection of chaoses appearing to be outside of the current extent of the SP plains. The blocks in this region resemble blocks in other regions with the exception of the more rounded appearance of many blocks (both bases and tops). Where the blocks do not appear rounded the blocks sometimes maintain the texture of the surrounding terrain outside SP, similar to blocks in other regions. | 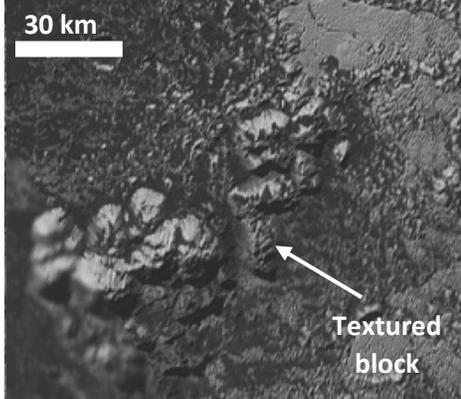 |

## 2.2 Europa

Chaotic terrain is one of the major terrain types found on Europa and may cover ~25–40% of Europa's surface (Riley et al., 2000; Figueredo and Greeley 2004; Collins and Nimmo, 2009). On Europa, chaotic terrain disrupts the pre-existing ridged and banded terrain to varying degrees. In some chaotic regions large blocks are left relatively intact and the ridged terrain preserved on the block tops can be used in some cases to infer the degree of translation, rotation, and tilting experienced by the blocks (Conamara Chaos in **Fig. 1**; e.g., Spaun et al., 1998; Collins and Nimmo, 2009). Other regions are broken down almost entirely into a rubbly matrix material (sometimes referred to as "rugged hummocky terrain"; Figueredo and Greeley, 2004) where no larger blocks are preserved. Here we map the preserved blocks in Conamara Chaos and in two other regions (described in **Fig. 1**). Conamara is a classic example where many large blocks are contained within a relatively distinct, scarp-bounded region. There are no other examples quite like Conamara in Europa images, but the two other regions are the closest comparisons that could be found.

Theories for the formation of europan chaotic terrain include a variety of subsurface processes which could cause the fracturing and, in some cases, the subsequent movement of chaotic blocks. Collins and Nimmo (2009) summarized five general models for the formation of chaotic terrain that have commonly been proposed in the literature that involves some variation of: (1) melting through the icy shell (Fagents, 2003; Goodman et al., 2004; Schmidt et al., 2011; Quick et al., 2017), (2) diapirism (Pappalardo et al., 1998; Pappalardo and Barr, 2004; Figueredo and Greeley, 2004; Nimmo and Giese, 2005), (3) brine mobilization (Head and Pappalardo, 1999; Han and Showman, 2005), (4) injection of sills (Collins et al., 2000; Michaut and Manga, 2014; Craft et al., 2016; Culha and Manga, 2016; Manga and Michaut, 2017; Noviello et al., 2019), and (5) impacts (Cox et al., 2008; Cox and Bauer, 2015). These processes range from those with only minor components of melted material (e.g., diapiric upwellings with partial melting, intrusions from the ocean below), to large-scale melt-through of the ice shell. Chaos morphologies vary greatly, and here we focus specifically on the morphology of selected regions where chaotic blocks can be distinguished from both the background terrain and rubbly chaotic matrix material.

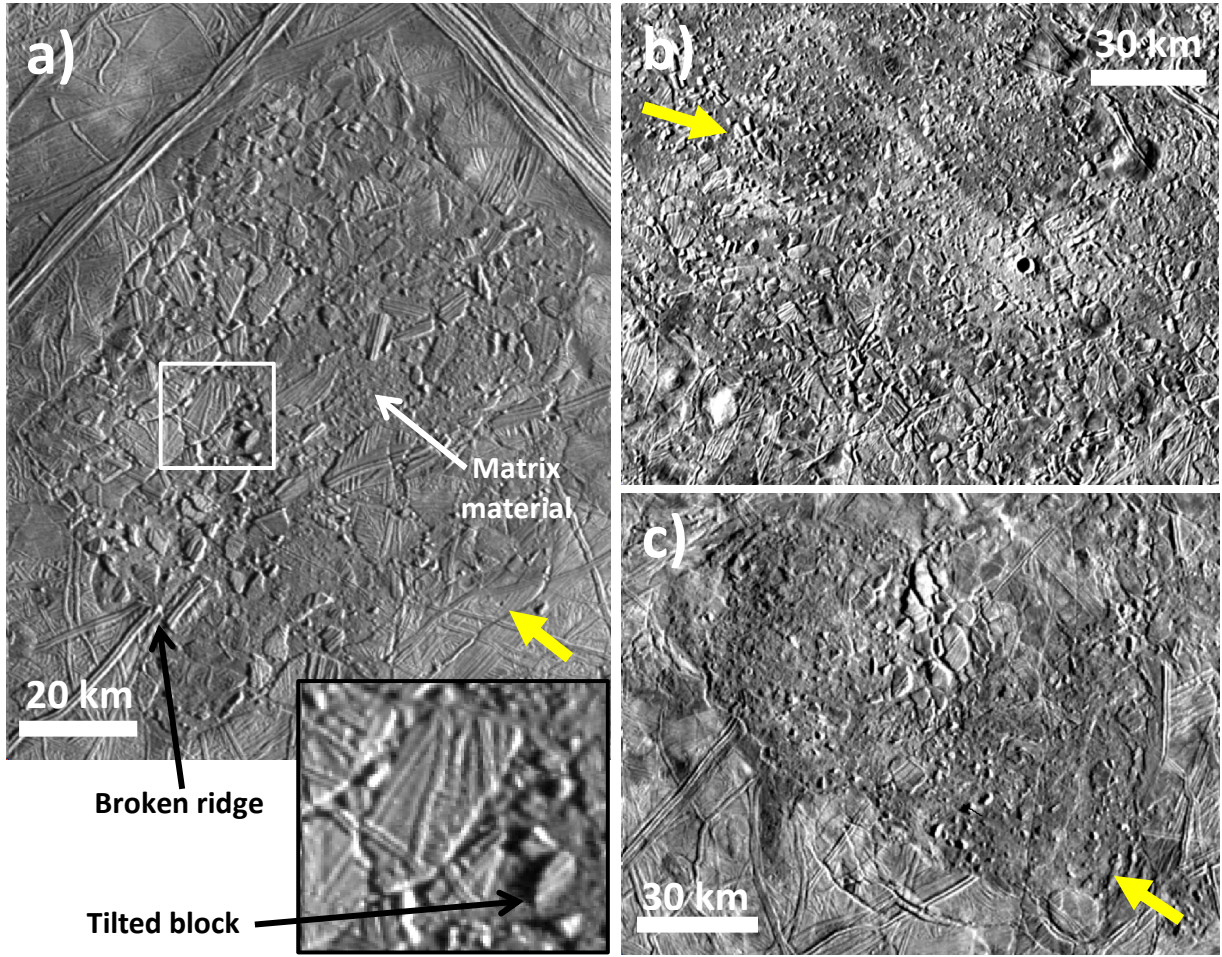

**Figure 1. Chaotic terrain regions on Europa.** a) Conamara chaos region (centered at 10° N, 274° W), b) East RegMap region (centered at 26° S, 83° W), and c) West RegMap region (centered at 0° N, 235° W). Note that the chaos regions in the East and West RegMaps are somewhat connected to the surrounding chaos compared to chaos in Conamara, which appears as a more isolated region. The chaotic terrain blocks sometimes appear to have been rotated and/or tilted to varying degrees (e.g., Spaun et al., 1998) in the surrounding broken up lumpy matrix material, which is also widely considered as chaotic terrain (Collins and Nimmo, 2009). Base mosaic resolutions: 180 m px$^{-1}$ and 210–220 m px$^{-1}$ for Conamara and RegMap regions, respectively (see Section 3.1.2 for further information). Global context of each mapping region is provided in Figure A2, Appendix A. Yellow arrows indicate approximate illumination direction. Note Europa has a west-positive longitude system (https://on.doi.gov/36k66To).

## 2.3 Mars

Chaotic and fretted terrains are both lowland terrain types in the equatorial and mid-northern latitudes of Mars (Carr, 2001). We studied chaotic terrain in the near-equatorial region east of Valles Marineris near the Xanthe and Margaritifer Terrae within the extent of 15° N, 300° W, and 19° S, 341° W, and fretted terrains along the dichotomy boundary in the Ismenius Lacus quadrangle between 25° N and 50° N and from 280° W to 350° W. General descriptions of chaotic and fretted terrain regions are provided in **Table 2** (see **Fig. A3** for an overview of mapped chaotic and fretted terrain areas on Mars). In addition to the close geographic relationship of these terrain types, some common characteristics can be observed between the two terrains. Both terrain types are dominated by large steep-sided mesas and smaller crest-topped knobs separated by flat smooth-floored valleys. Both terrain types are also situated in areas with evidence for past high rates of aqueous activity, as outflow channels have incised and carved the landscape where the majority of mountain blocks are found (e.g., Irwin et al., 2004; Rodriquez et al., 2005). In addition, fracturing of the surrounding highland terrain is commonly observed in the areas where both terrain types are found and has likely played a key role in the formation of both terrains. The size, orientation, and spacing of these extensional fractures are likely directly correlated to the size and morphology of chaotic terrain blocks within each region (Rodriquez et al., 2005; Warner et al., 2011). The extensive fracture systems in the highlands have been proposed to be produced by extensional stresses related to subsidence (Rodriguez et al., 2005). The similarities between the morphology of the terrain types could imply that some of the same geomorphic processes have shaped both terrains. Martian chaotic terrains have previously been compared to submarine polygonal slope failures on Earth (e.g., Rossi et al., 2006; Moscardelli et al., 2012), which has been noted to bear resemblance to polygonal faulting on venusian plains (e.g., Hamilton, 2015). By including fretted terrain in our comparison of chaotic terrains we seek to quantify any morphological similarity between martian chaos and fretted terrain.

A variety of formation theories have been proposed for martian chaos. The most common formation scenarios discussed include: (1) catastrophic or non-catastrophic release of groundwater stored within confined aquifers beneath the cryosphere or extensive subterraneous caverns (Carr, 1979; Lucchitta et al., 1994; Chapman et al., 2003; Ogawa et al., 2003; Rodriguez et al., 2003, 2005; Leask et al., 2006;), (2) catastrophic formation and transport of water- and/or carbon dioxide-charged debris flows (Nummedal and Prior, 1981; MacKinnon and Tanaka, 1989; Tanaka, 1997, 1999; Hoffman, 2000; Tanaka et al., 2001), (3) magma-ice interactions in the near cryosphere (Masursky et al., 1977; Chapman and Tanaka, 2002; Head and Wilson, 2002; Chapman et al., 2003; Leask et al., 2006; Meresse et al., 2008), and (4) release of gas or gas-water mixtures from liquid and solid carbon dioxide and/or decomposed gas hydrates (Milton, 1974; Lambert and Chamberlain, 1978; Hoffman, 2000; Komatsu et al., 2000; Max and Clifford, 2001; Tanaka et al., 2001). Variations in the geomorphology (see **Table 2**) of different chaos regions and individual characteristics of chaos terrain between different regions imply that multiple formation mechanisms may have operated (Warner et al., 2011). Studies have suggested that the development of fretted terrain could have been initiated by fracturing (Irwin et al., 2004) or through some variation of fluvial (Breed et al., 1982; Carr, 2001), glacial (Lucchitta, 1984; Head et al., 2010; Sinha and Murty, 2013), and/or mass wasting erosive processes (Squyres, 1978; Carr, 1995; Carr, 2001). The fretted terrain in this region is thought to have developed during the Early Hesperian Epoch (Irwin and Watters, 2010; Tanaka et al., 2014), which ties the formation of fretted terrains close to the formation of chaotic terrains thought to have formed generally in the Late Hesperian (Rodriguez et al., 2005; Meresse et al., 2008; Tanaka et al., 2014).

**Table 2. General summary of regions studied on Mars.** The studied chaotic terrain regions are grouped together according to the predominant type of block morphology. The pixel scale for all images in this table is 100 m px$^{-1}$ (see Section 3.1 for information about image datasets).

| Type* | Regions | General Description | Image Example(s) of Chaotic Terrain Type** |
|---|---|---|---|
| "Mesa shaped" chaotic terrain | *Aram Chaos* (3° N, 338° E) *Chryse Chaos* (12° N, 323° E) *Hydraotes Chaos* (1° N, 325° E) *Hydraspis Chaos* (4° N, 331° E) *Iani Chaos* (1° S, 342° E) | The chaos in these regions are dominated by large flat-topped mesas and smaller knobby blocks separated by narrow valleys. The tops of the large mesas roughly match the elevation of the surrounding plateau. Smaller knobs with relatively lower height to diameter ratios compared to mesa blocks can be found intermixed between and around the mesas. Fracturing of the surrounding plateau into blocks is commonly observed throughout many regions. | Hydraotes Chaos; 60 km; Mesa |
| "Knobby" chaotic terrain | *Arsinoes Chaos* (8° S, 332° E) *Aureum Chaos* (4° S, 333° E) *Aurorae Chaos* (9° S, 324° E) *Eos Chaos* (17° S, 315° E) *Pyrrhae Chaos* (10° S, 331° E) | The chaos in these regions are dominated by "knobby" shaped blocks, and a smaller distribution of mesa shaped blocks that are usually found near the edges of the chaotic terrain basins where fracturing of the surrounding plateau is observed. The knobby blocks typically have angular or sub-rounded "peaks", but sometimes feature roughly flat tops. The blocks are separated to varying degrees by the smooth channel floor. | Aurorae Chaos; 100 km; Knob |
| "Smaller" chaotic terrain | *Iamuna/ Oxia Chaos* (0° N, 320° E)<br><br>*Ister Chaos* (12° N, 303° E) | The chaos in these regions is located within patches of subcircular depressions on the channel floor of outflow channels. The blocks generally appear as smaller mesas (compared to mesas found in other regions) with flat tops that match the elevation of the surrounding channel bed, and sometimes maintain the surrounding flow pattern. The fractures in these regions are interpreted to not have been significantly deepened by erosion. | 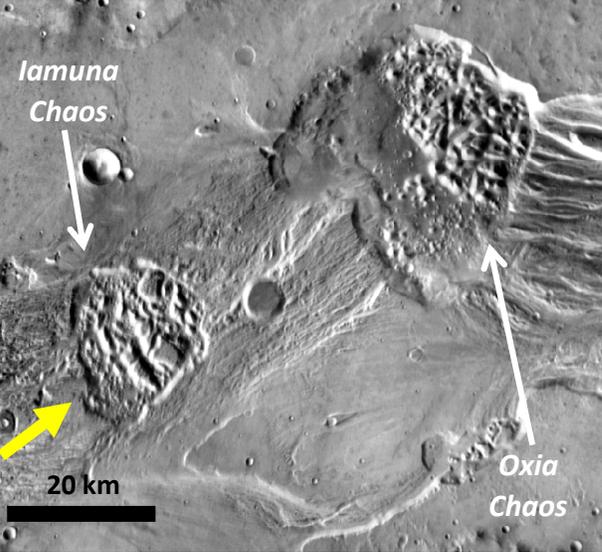 |
| Lowland fretted terrain | *Ismenius Lacus quadrangle* (44° N, 26° E) | The region features large steep-sided mesas, plateaus and knobby blocks separated by broad valleys that are significantly larger than the mountain blocks in our studied chaotic terrain regions. The tops of the blocks roughly maintain the same elevation as the surrounding plateau but become visibly shorter in the northern extent of the region through what appears to be an erosional contrast marked by darker deposits. Moving northwards, the blocks gradually become shorter and eventually merges into the surrounding terrain. | 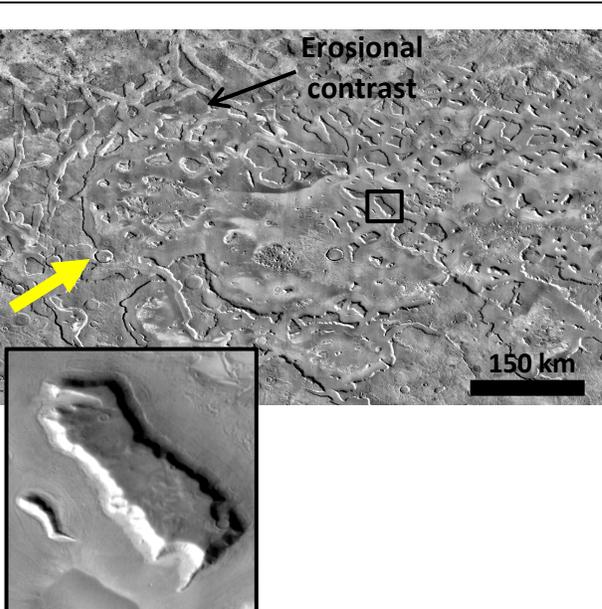 |

*Regions are grouped into categories based on the most common block morphology from visual inspection.
**The image examples are selected to provide the best visual example of the chaotic terrain type described. Yellow arrows indicate approximate angle of illumination.

# 3. MAPPING METHODS
## 3.1 Image and Elevation Datasets
### 3.1.1 Pluto

The *New Horizons* mission flew through the Pluto system in July of 2015, and collected various datasets including visual-band, color, and spectral imaging of Pluto's surface and

11 / 45

atmosphere (e.g., Stern et al., 2015). Medium- and high-resolution images covering SP allowed us to study the chaotic mountain blocks found within it. Mapping was conducted throughout five regions using a wider, ~315 m px$^{-1}$ observation covering all of the mountain ranges, and also on several image strips at better resolution (77–125 m px$^{-1}$) covering a portion of the mountains (Singer et al., 2021; see their Fig. S1 and Table S1 for image product details). A 240 m px$^{-1}$ stereo digital elevation model (DEM) was utilized to guide mapping throughout all regions (Schenk et al., 2018), and derive block heights.

### *3.1.2 Europa*

The *Galileo* mission orbited Jupiter for almost eight years (1995–2003) and made close passes to its major moons including Europa. We used the mid-resolution regional mapping east and west longitudinal strips at 210–220 m px$^{-1}$ (referred to as the East and West RegMaps). The DEM is an albedo-controlled photoclinometry product with the same resolution as the base mosaics (see methods in Schenk, 2002; Schenk et al., 2004; Singer et al., 2012; Singer et al., 2018; Singer et al., 2021). For the region covering Conamara Chaos we used a base mosaic and DEM at 180 m px$^{-1}$. Vertical errors are discussed in Singer et al. (2021), but are estimated to be ~4 to 7 m per pixel at RegMap resolutions and ~3 to 6 m for the Conamara DEM. Random errors accumulate across the width of a feature, and for uniform photometric properties, these errors should be on the order of $\sqrt{n}$ x the pixel height error, where n is the pixel scale of the feature of interest.

### *3.1.3 Mars*

We used the *Mars Odyssey* Thermal Emission Imaging System (THEMIS) daytime infrared global mosaic at 100 m px$^{-1}$, with a horizontal positional accuracy of ±100 m (Edwards et al., 2011). Topography data was from a 200 m px$^{-1}$ product that combines data from NASA's *Mars Global Surveyor* Mars Orbiter Laser Altimeter (MOLA) and the European Space Agency's *Mars Express* High-Resolution Stereo Camera (HRSC), with a total elevation accuracy of ±3 m (Fergason et al., 2018). Both image and topography were used to map both chaotic and fretted terrain regions, and are publicly available for download through USGS Astrogeology's websites (https://on.doi.gov/2plQPR4 and https://on.doi.gov/2pjXwmH, respectively).

## 3.2 Mapping Considerations

Chaotic terrain blocks were mapped in ArcGIS using polygons to outline the perimeter of each block along their apparent base, using visual identification and topographic mapping for confirmation. For some regions of chaotic terrain defining the boundary between mountain blocks is a relatively simple task because the flatter surrounding terrain clearly intersected with the steeper block walls with a sharp transition (e.g., martian blocks). However, in other regions the separation between blocks was unclear because the blocks abut each other, or the matrix material (or chaotic inter-block material) was also somewhat blocky and at a similar elevation to the large blocks (e.g., **Fig. 1**). Topographic information was helpful for distinguishing blocks from each other and the surrounding terrain in these cases. Blocks were not included into our measurements if their boundaries were indeterminable by eye or with topography, or in the case where they appear to have not been completely fractured from the surrounding topography (**Table 2**; e.g., Aram Chaos). Blocks below a given diameter for each region were excluded from our measurements due to resolution and terrain constrains, and to improve our measurement accuracy and feature identification. There are blocks smaller than our diameter cutoff across all regions studied, and therefore implementing a cutoff artificially truncates our distribution. These results and conclusions are described with this information in mind.

On Pluto mountain blocks were not analyzed below a diameter of 3 km due to the resolution and terrain constrains. Below this cutoff blocks were hard to define due to their low degree of separation, and in many cases merged with the surrounding chaotic inter-block material or other surface features found across the mountain ranges (see White et al., 2017 for geologic mapping of SP). Similarly, some of the smaller blocks on Europa were harder to distinguish due to the rubbly matrix material or pieces of ridges that can sometimes resemble a block (see **Fig. 1a**). We focused on the blocks that were more easily distinguished because their boundaries were expressed by distinct shadows and contrasts of the ridged block top with the surrounding rubbly matrix (**Fig. 1**). Because europan blocks below the size of 2 km were difficult to accurately measure, blocks below this cutoff were not included in our mapping efforts. In a few cases, the blocks found in regions on Europa appeared to have been slightly tilted and rotated even if their boundaries were distinguishable. Because tilting of blocks could influence our measurements of height and size, blocks on Europa were labeled to indicate whether they appeared tilted or not (**Fig. 5b**). For Mars, (chaotic and fretted terrain) blocks were excluded from our measurements below a diameter of 3 km in order to improve feature identification. Below this cutoff blocks could resemble other topographic features (e.g., volcanic cones; Meresse et al., 2008), and were hard to define due to their sometimes lower degree of separation (e.g., smaller mountain blocks in Aureum Chaos and Aurorae Chaos). Blocks below this size were, however, included in our mapping efforts for a few regions (Iamuna/Oxia Chaos and Ister Chaos) because the majority of blocks in these small areas were below our diameter cutoff and were distinguishable from other features. For Mars, we also noted whether the blocks were flat-topped (appearing to have flat tops that represented a pre-existing surface similar to the elevation of the surrounding terrain) or crest-topped (blocks with no apparent pre-existing surface and that form a sharp peak or ridge at the top). Although there are a few blocks that do not fall strictly into these two morphologic categories, these were by far the most predominant types (see **Fig. 5c**).

A final consideration to note is the systematic effects posed by the variability of the robustness of mission data sets available for each body (e.g., availability of multiple viewing and lighting geometries, and degree of spatial coverage and resolution). We minimized the systematic effects associated with our measurements across the three bodies by incorporating the different mapping and resolution cutoffs described in this section.

### 3.3 Block Measurements

To derive a measure of the mountain block size (diameter) we used the surface area of the block (measured geodetically on a sphere) and calculated an equivalent diameter as if the feature were a circle (i.e., using Area$_{circle}$ = $\pi r^2$ to solve for diameter). The uncertainties in the block diameters are more dependent on decisions of where the base of the block is, rather than any systematic items such as pixel scale. We have described the image constraints and geologic variability that factor into decisions about mapping in **Sections 3.1** and **3.2**, but we attempted to maintain a consistent mapping strategy on each body.

The height of each block was estimated by subtracting the average elevation at the base of the block (basal perimeter of the mapped feature outline) from the maximum elevation on the block top (maximum elevation point within each polygon). The uncertainty in the calculated height is mainly influenced by variations in the terrain (see **Section 3.1** for discussion of vertical errors). **Appendix B** presents the uncertainty in the estimated block height for each region, which ranges from a few meters up to 10–20 m (see **Appendix B** for further information). There are some points about the height measurements that should be noted: (i) in some cases we cannot tell if or how far the block extends under the surrounding terrain (e.g., for the Pluto blocks in the nitrogen-rich ice sheet of SP), but in each case we measure the height of the block that is visible, and (ii) in some cases the blocks are closely joined without a large degree of separation, which can lead to measuring the base of the block higher than the true base.

The axial ratio (long axis divided by short axis) was determined by using the minimum bounding geometry rectangle of the smallest area enclosing the block.

# 4. RESULTS

## 4.1 Diameter Distributions

A summary of the number and average characteristics (diameter, height, and axial ratio) of mapped mountain blocks is given in **Table 3**. For all mountain blocks combined on Pluto we find effective diameters to be approximately log-normally distributed, with a peak between 5 and 13 km (**Fig. 2a**). Keeping in mind the lower-limit diameter cutoff is ~3 km for Pluto, this peak represents the most typical size above 3 km. Some regional variation between the mapped mountain ranges can be observed, as the size distribution for blocks in Tenzing Montes appear to be slightly shifted towards larger block sizes with a broader peak from ~8 to 18 km (**Fig. 2a**). Although Zheng He Montes consists of only about four distinct mountain blocks, this region features some of the largest mapped blocks measuring between ~21 to 37 km in size (**Fig. 2a**).

Measured chaotic blocks on Europa across all regions show a peak distribution between 2 and 5 km in size (**Fig. 2b**) with little variation in the effective diameters between each location. In this case, however, the lower-limit diameter cutoff (3 km) intersects the lower edge of this peak, thus it is not clear if smaller blocks would be even more frequent given higher resolution images. Blocks in the West RegMap region are slightly shifted towards smaller block sizes, but generally follow a similar size trend as blocks in Conamara and East RegMap. Our measurements show that blocks in europan chaos typically do not exceed above a maximum of ~14 km in size, and only rarely exceed above 6–8 km in diameter (**Fig. 2b**).

On Mars chaotic blocks displayed a wide distribution in block sizes. Blocks in most regions have peak distributions near 3–4 km (**Fig. 2c**). **Figure 2c** shows that some regional variation in the size of martian mountain blocks can be observed, as blocks in Aram Chaos, Aurorae Chaos, Chryse Chaos, Hydraotes Chaos, Eos Chaos, and Iani Chaos are shifted towards larger block sizes. Blocks were mapped to the approximate diameter cutoff—because it is not clear exactly which bocks will be under the cutoff diameter when mapping, we mapped some under the cutoff in order to be more complete down to the cutoff diameter. Martian chaos blocks typically range up to a maximum size of 10 km (**Fig. 2c**) and are less frequently observed above this size. Our measurements show that fretted terrain blocks are significantly larger in size than chaos terrain blocks. Fretted terrain blocks have a broader peak distribution between ~4 to 9 km and are frequently observed up to 10–20 km in size (**Fig. 2c**) but can range up to above 70 km in size (**Fig. 5c**).

**Table 3. Summary of number of mapped blocks, and averages of block characteristics.**

| Body | Region | Number of Mapped Blocks | Average Block Height (km) | Average Block Diameter (km) | Average Axial Ratio of Blocks |
|---|---|---|---|---|---|
| *Mars* | *All chaos regions** | *8027* | *0.47* | *4.40* | *1.41* |
| | *Aram Chaos* | 173 | 0.48 | 6.41 | 1.59 |
| | *Arsinoes Chaos* | 293 | 0.46 | 3.88 | 1.46 |
| | *Aureum Chaos* | 822 | 0.48 | 4.06 | 1.39 |
| | *Aurorae Chaos* | 2158 | 0.45 | 4.58 | 1.40 |
| | *Chryse Chaos* | 1202 | 0.42 | 4.03 | 1.43 |
| | *Eos Chaos* | 847 | 0.63 | 5.00 | 1.39 |
| | *Hydraotes Chaos* | 1061 | 0.58 | 5.20 | 1.45 |
| | *Hydrapsis Chaos* | 831 | 0.32 | 3.01 | 1.36 |
| | *Iamuna/Oxia Chaos* | 57 | 0.21 | 2.30 | 1.47 |
| | *Iani Chaos* | 426 | 0.46 | 4.83 | 1.42 |
| | *Ister Chaos* | 95 | 0.15 | 2.68 | 1.46 |
| | *Pyrrhae Chaos* | 62 | 0.48 | 5.09 | 1.61 |
| | ***Fretted Terrain*** | ***1676*** | ***0.72*** | ***9.84*** | ***1.62*** |
| *Pluto* | *All regions* | *378* | *1.18* | *10.4* | *1.44* |
| | *Al-Idrisi Montes* | 157 | 0.87 | 9.70 | 1.55 |
| | *Zheng He Montes* | 4 | 3.77 | 26.8 | 1.51 |
| | *Baret Montes* | 45 | 0.83 | 8.93 | 1.36 |
| | *Hillary Montes* | 94 | 1.37 | 9.86 | 1.37 |
| | *Tenzing Montes* | 61 | 1.69 | 12.5 | 1.36 |
| | *Tabei Montes* | 17 | 1.51 | 11.4 | 1.36 |
| *Europa* | *All regions* | *401* | *0.16* | *4.28* | *1.59* |
| | *Conamara* | 139 | 0.13 | 4.82 | 1.70 |
| | *West RegMap* | 106 | 0.21 | 3.91 | 1.61 |
| | *East RegMap* | 156 | 0.14 | 4.11 | 1.45 |

*Fretted terrain statistics are provided separately below and are not included in this line.

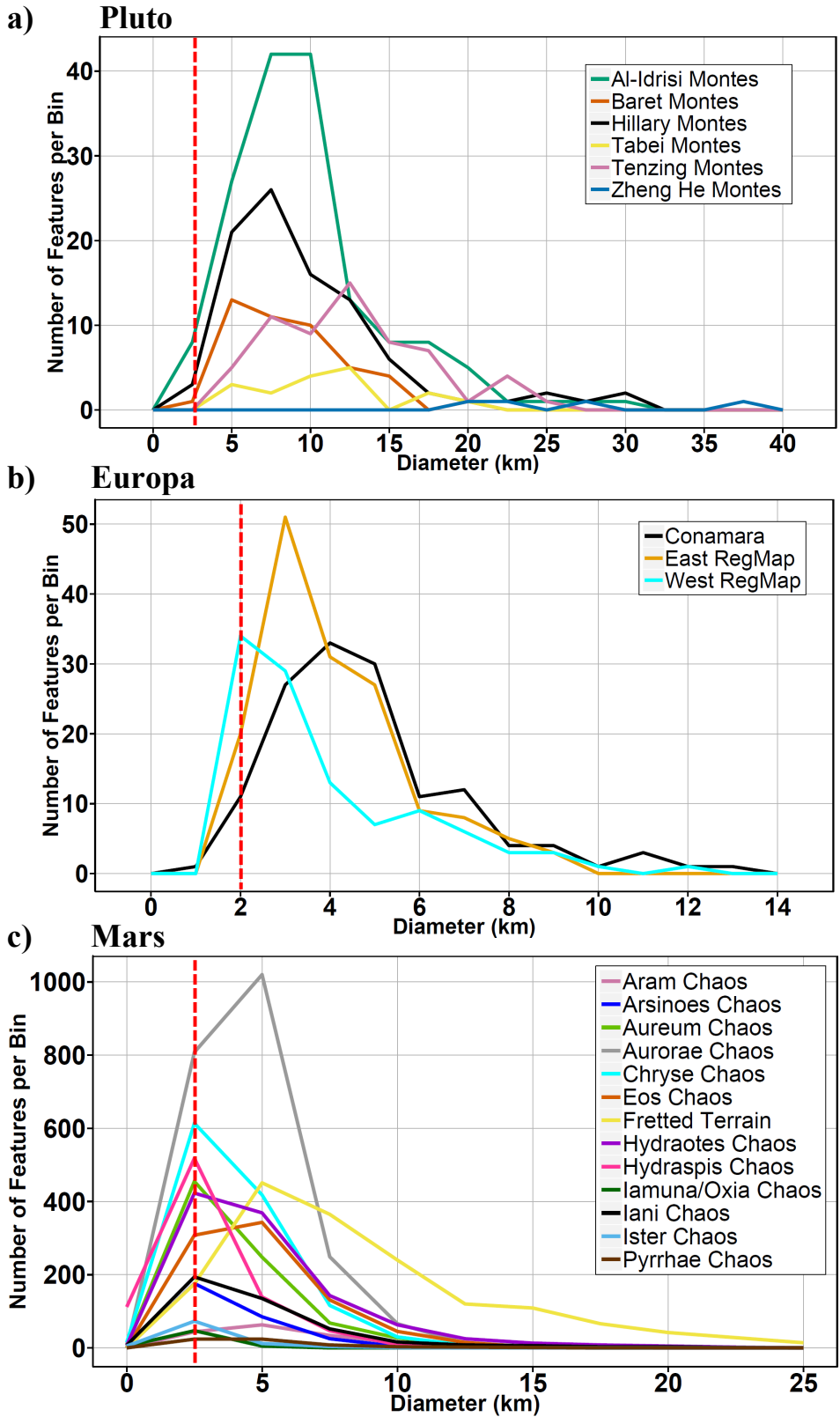

**Figure 2. Histograms of effective feature diameters.** Shown for a) size distributions of chaotic mountain blocks across studied regions on Pluto. Mountain blocks on Pluto have a peak distribution in size between 5 and 13 km for all mapped mountain ranges. Mountain blocks in Tenzing Montes appear to be shifted towards slightly larger blocks, and features a broader peak distribution between 8 and 15 km. b) Europan chaos blocks displayed a peak distribution between 2 and 5 km in size, with little regional size variation. c) Mountain blocks on Mars typically measured up to 5 km in size, with some regional variation in respective peak size distributions. Respective mapping cutoffs are denoted by dashed red vertical lines. Fretted terrain blocks are significantly larger than chaos blocks, and appear to be shifted towards significantly larger block sizes with a peak distribution between ~5 and 10 km. Note that the axes ranges are different for each body.

## 4.2 Height Distributions

Mountain blocks on Pluto most commonly measured between 0.5 and 1.5 km in height (**Fig. 3a**). Some regional variation in the height of mountain blocks on Pluto can be observed. Mountain blocks in Al-Idrisi Montes and Baret Montes display a peak height distribution between 0.5–1 km (**Fig. 3a**), and 1–1.5 km for blocks in Tenzing Montes, Hillary Montes, and Tabei Montes (**Fig. 3a**). Blocks in Zheng He Montes does not display much variation in height, and the four blocks within the region measures between ~3.5–4 km in height. Tilting of blocks could influence the apparent height of blocks on Pluto. We measured the slopes of blocks with dipping planes, or "tops", that resembles the texture of the surrounding unfractured terrain. The slope of the tops of blocks measured between 3–9° for mountain blocks in Al-Idrisi Montes, 7–13° for mountain blocks in Zheng He Montes, and 4–10° for mountain blocks in Tenzing Montes. Because these slopes typically measured <10° and most blocks do not appear obviously tilted, tilting likely only plays a small role on our height estimates for blocks on Pluto. Blocks in Baret Montes and Tabei Montes generally do not appear significantly tilted, and therefore we did not measure the slopes of blocks studied in these regions. Individual slope values for each measured block is provided in **Table C1** (see **Appendix C**). The slopes of the fracture face, or block side, measured between 13–27° for blocks in Al-Idrisi Montes, and 30–34° and 28–32° for mountain blocks in Zheng He Montes and Tenzing Montes, respectively.

The height distribution of europan blocks is significantly more homogenous compared to blocks on the two other bodies. Block height measurements on Europa show that the typical height of blocks is between 0.1 to 0.2 km (**Fig. 3b**). A small number of blocks exceed a height of 0.2 km, and some blocks in West RegMap reach up to slightly above 0.5 km in height (**Fig. 3b**). A brief discussion of the implications for tilted blocks is included in **Section 4.4** below, and blocks are categorized as tilted (or not) in **Figure 5b**.

Our measurement of martian blocks across chaotic and fretted terrain regions show that there is some regional variation in the height of blocks across both terrain types. Nearly all chaos regions (including fretted terrain) showed a peak in the height distribution around 0.5 km (**Fig. 3c**), although martian chaos and fretted terrain blocks can measure up to around 2.5–2.8 km tall. Blocks in most mapped regions rarely exceed ~1 km in height (**Fig. 3c**), but mountain blocks in fretted terrain and chaotic terrain regions Aureum Chaos, Aurorae Chaos, Hydraotes Chaos, and Eos Chaos have a higher frequency of blocks reaching above 1.5 km in height (**Fig. 3c**). The blocks within Iamuna/Oxia Chaos, and Ister Chaos regions are the shortest and have peak distributions below 0.5 km (**Fig. 3c**). Mountain blocks in fretted terrain follow the same height distribution as chaos blocks but are slightly shifted towards taller block sizes. Trends in height with latitude for fretted terrain blocks are discussed in **Appendix D**.

The gravity on each of these bodies differs and the material strength also likely differs. The surface gravities on the two icy worlds Pluto and Europa are 0.62 and 1.35 m s$^{-2}$, respectively. Mars has the highest surface gravity (3.71 m s$^{-2}$). The material strengths are not well known for the regions studied here, as the strengths will depend on many factors such as compositions, geologic histories, and temperatures. In a simple comparison, for an equal-strength material, Pluto could support a topographic load placed on the surface of about ~2 times higher than on Europa, and ~6 times as higher than that on Mars. The maximum block sizes on each body do not follow these rules, and the features themselves are not necessarily simple loads placed on a surface layer (rather they are likely the breakup of an existing layer), so this trend is not expected. Additionally, the chaotic blocks studied here are not the tallest features on each body, so this is additional evidence that their heights are likely not strength-limited.

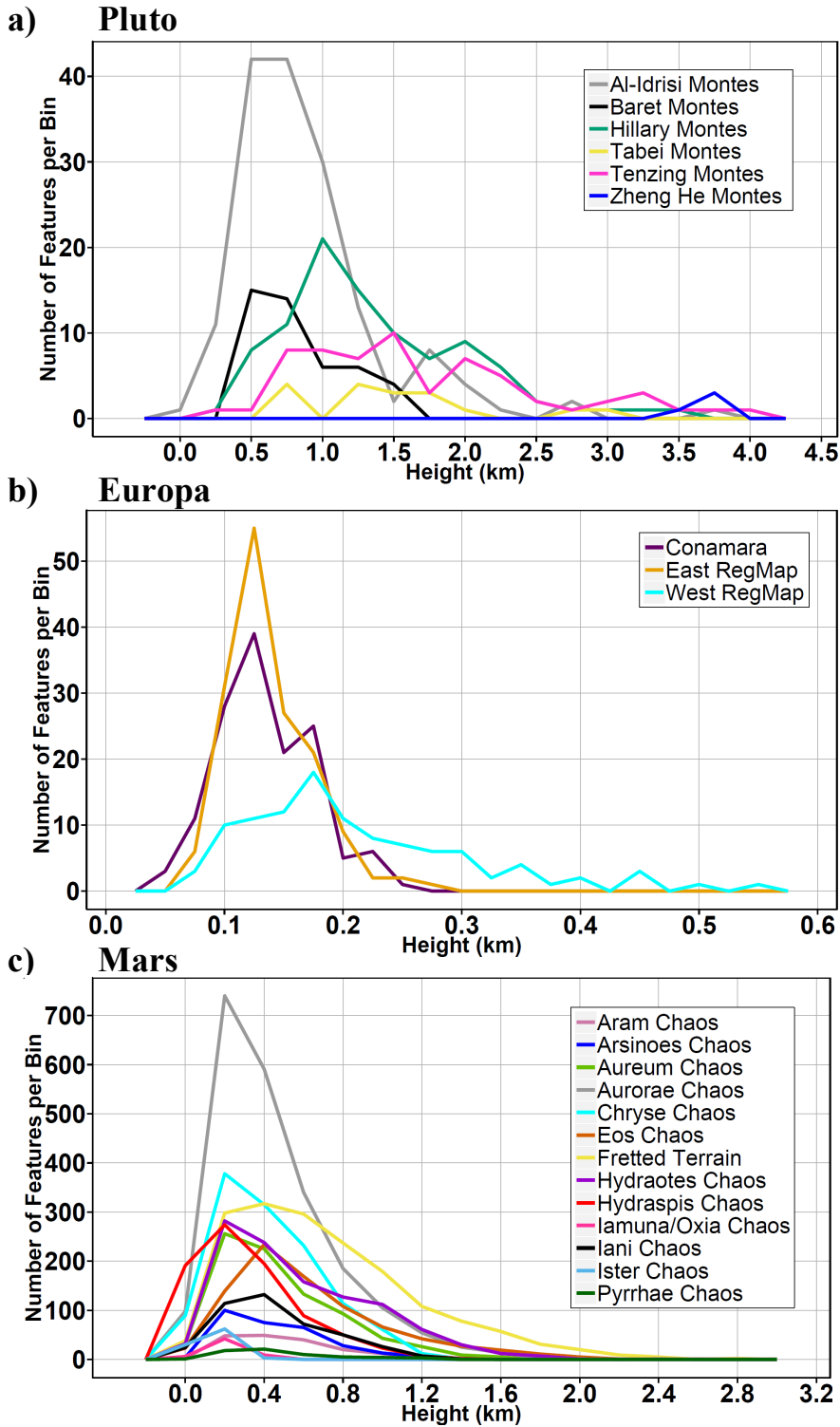

**Figure 3. Histograms of feature topographic signatures.** Shown for height of blocks within mapped locations on a) Pluto, b) Europa (see Fig. 5b for frequency of tilted blocks in each region), and c) Mars. Note that the axes ranges are different for each body.

## 4.3 Axial Ratio Distributions

The axial ratio of the blocks studied generally measured between 1 and 2 for all studied regions across all bodies (**Fig. 4a–c**). Mountain blocks on Pluto showed little variation in axial ratios across the regions studied, with a peak distribution around 1 (**Fig. 4a**). On Europa blocks had a peak distribution between ~1.25–1.75 (**Fig. 4b**). Blocks in Conamara Chaos appear to have a slightly broader range of axial ratio values but follow similar axial ratio trends as blocks in the East and West RegMap. Martian blocks have typical axial ratios between 1 and 1.5 (**Fig. 4c**), and fretted terrain blocks had a slightly higher average axial ratio compared to chaotic blocks. Generally, fretted terrain blocks appear as large mesas and plateaus that have a clear long axis (that generally appear parallel to the dichotomy boundary) compared to the more square blocks in martian chaos (**Table 2**). It is possible the stress field that created fractures along the dichotomy boundary was somewhat different than that for other chaotic terrains. Due to the low variation of axial ratios for blocks across the three bodies, we do not extend these observations into our discussion (**Section 5**).

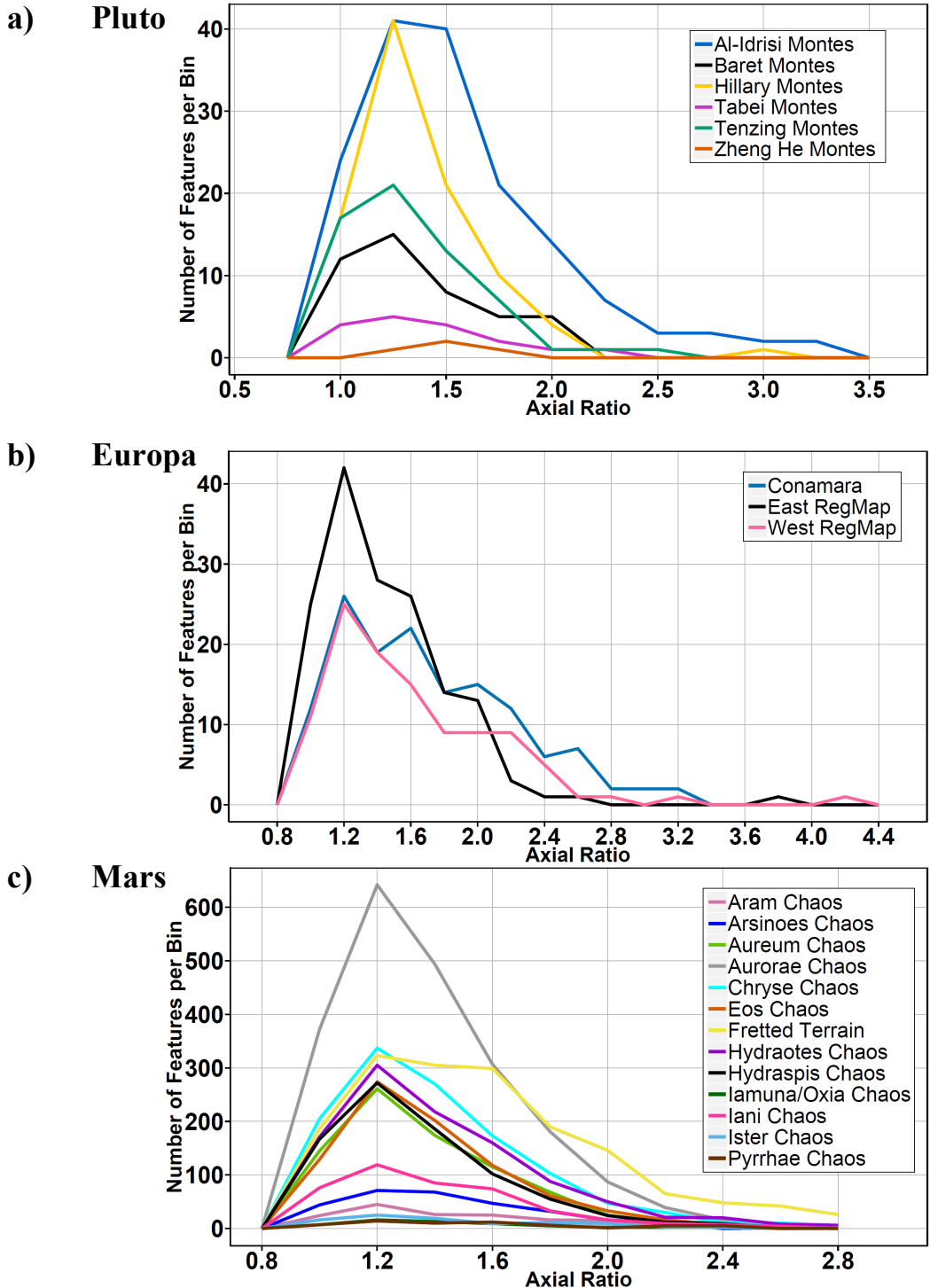

**Figure 4. Histogram of axial ratio signatures.** Shown for calculated axial ratios for mapped regions across a) Pluto, b) Europa, and c) Mars. Most axial ratios for all mapped blocks ranged between 1 and 2, with some minor regional variation. Note that the axes ranges are different for each body.

## 4.4 Trends in Feature Topography with Diameter

The size and height distributions for regions across all three mapped regions are presented in **Figure 5a–c**. For chaotic terrain regions on Pluto and Mars we find a positive linear relationship between height and diameter **(Figs. 5a, c)**. Mountain blocks on Pluto reach up to ~4 km tall and ~38 km wide, and some regional variation can be seen between the mountain ranges. Blocks in Tenzing Montes and Hillary Montes generally appear taller for the same effective diameter compared to other regions. This trend is consistent with the spikier appearance of the blocks in this region (see **Table 1**), as discussed in **Section 2.1**. Blocks in Al-Idrisi Montes (**Fig. 5a**) are on average shorter for the same effective diameter compared to the other ranges. This trend could be a result of the large variations in $N_2$ ice loss and accumulation in the northern edge of SP (Bertrand et al., 2018; their Fig. 10 and 14) suggesting stronger erosion of Al-Idrisi Montes blocks, which in this region do appear more eroded (fewer sharp edges as seen in **Table 1**).

The chaotic blocks on Europa exhibit a "flat" trend, as block height does not generally increase with increasing block size (**Fig. 5b**). The typical height of blocks on Europa is between 0.1 to 0.2 km, however, untilted blocks in the West RegMap region features slightly taller blocks on average, reaching up to ~0.5 km in height. Some but not all of the tallest blocks also appear tilted, so it is not clear if tilting alone could account for all the height variation between regions. In all regions on Europa, the proportion of tilted blocks is higher for smaller blocks (**Fig. 5b**).

While there is a general trend for increasing height with increasing block diameter of both martian chaotic and fretted blocks, there is also some regional variation in the size-height relationship of blocks across both terrain types. In certain regions, blocks generally do not exceed 1–1.5 km in height (e.g., Aram Chaos, Arisones Chaos, and Hydraspis Chaos), compared to blocks in other regions that reach up to above 2 km tall (e.g., Aurorae Chaos, Eos Chaos, and Hydraotes Chaos). Blocks in fretted terrain follow a similar general trend as blocks in chaotic terrains (**Figs., 5c, 6**) but some are significantly larger in size. The largest blocks (above ~30 km in size) do not show a strong trend, instead there is a range of heights from shorter to taller. A combined view of trends in feature topography with diameter for each region studied on Mars is provided in **Fig. E1** (**Appendix E**).

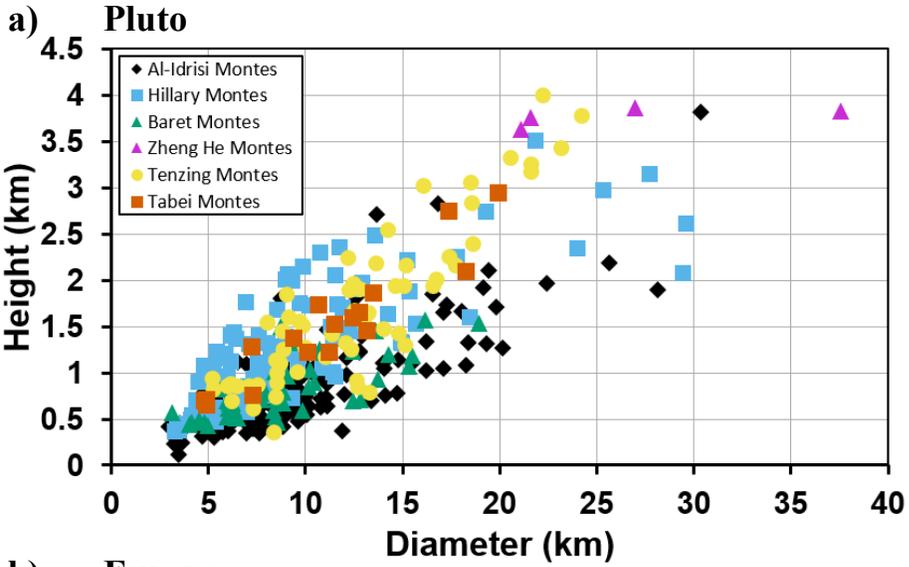
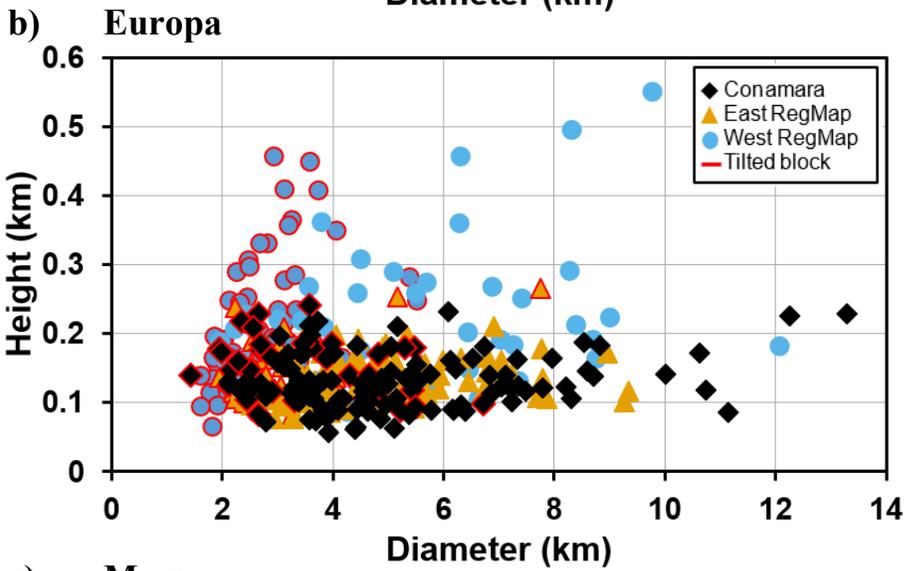
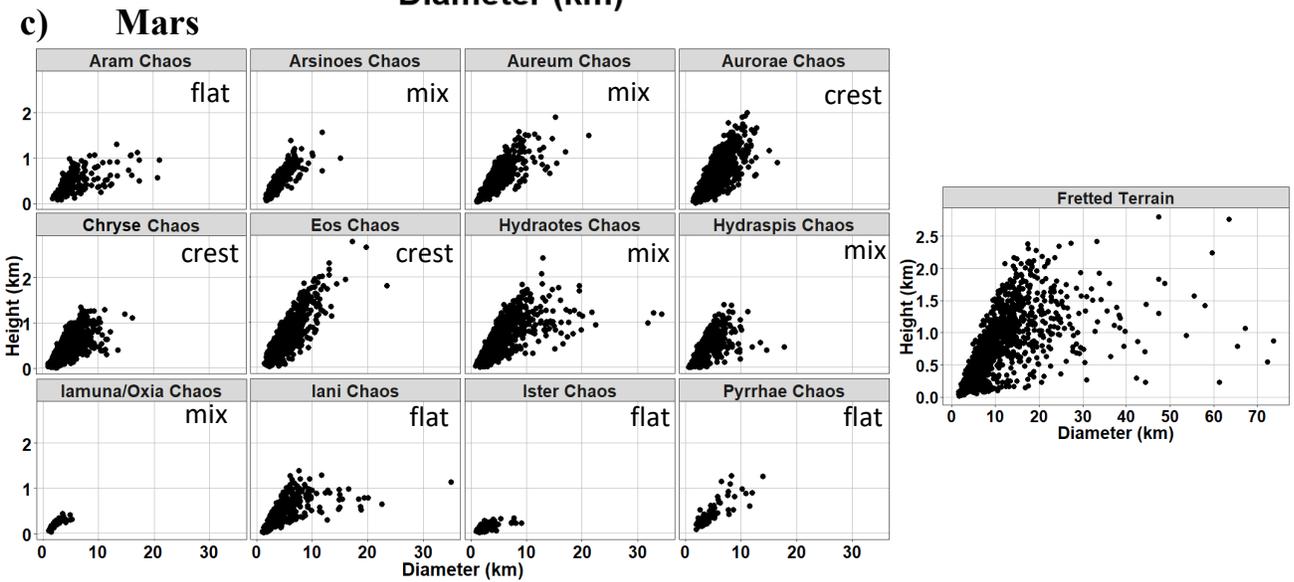

**Figure 5. Trends in feature topography with diameter.** Shown for a) Pluto, b) Europa, and c) chaotic and fretted terrain on Mars. For martian chaos regions, we also note the most common block top shape (e.g., more flat-topped blocks, more crest-topped blocks, or a relatively even mix of the two types). Mountain blocks on Pluto and Mars show a positive linear relationship between height and diameter, whereas blocks in europan chaos show a "flat" relationship. Note that the axes ranges are different for each body and terrain type.

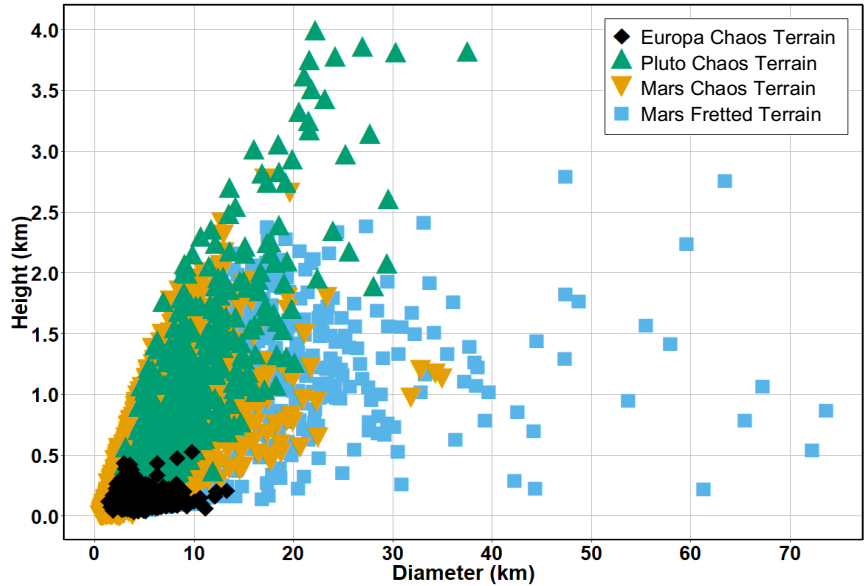

**Figure 6. Comparison between feature topography with diameter across the three studied bodies.** This combination plot shows the morphometric similarities between Pluto and Mars chaotic blocks although they are not necessarily formed and modified by the same processes (see discussion). Europan blocks are significantly shorter and show a flat trend. Martian fretted terrain blocks can be significantly larger compared with those in chaotic terrain regions on Mars and these larger blocks also appear to have a flatter trend.

## 5. DISCUSSION: IMPLICATIONS FOR CRUSTAL LITHOLOGY

The size and height distributions of chaotic mountain blocks could provide information about the lithologic structure and thickness of the fractured units on each body (**Fig. 7**). If the blocks are all the same height or reach a maximum height and level out (i.e. cease to increase in height with increasing diameter), then this could yield information about the layer thickness of the fractured unit. We speak first here about Europa because it provides an example of a distribution where the blocks do not show increasing heights with increasing size, unlike the mountain blocks on Pluto (see discussion below) and Mars (**Fig. 5**). Previous studies (e.g., Williams and Greeley, 1998) have used block heights in Conamara Chaos to estimate a 0.2–3 km thickness of the icy lithosphere on Europa using buoyancy models for floating iceberg-like blocks assuming the blocks were floating and reached an isostatic level (presumably during or shortly after chaos formation). The shell thickness is given by $\rho_s h/\Delta\rho$, where $\rho_s$ is the density of the substrate (liquid or solid the blocks are in), $h$ is block height, and $\Delta\rho$ is the density contrast between the solid shell and the substrate. This analysis using our typical untilted blocks heights on Europa of ~0.1 to 0.2 km, and less extreme compositions (Singer et al., 2021) leads to ice shell thickness predictions of ~1 to 4 km (**Fig. 8**) at the time of formation. For the taller untilted blocks around 0.5 km in the West RegMap region (**Fig. 5b**) the same analysis yields a slightly thicker lithospheric estimate of ~5 to at least 9 km (**Fig. 8**). This isostatic analysis applies if the blocks have maintained their isostatic positions over time, and any viscous relaxation of the block positions or elastic support would again imply these shell thickness estimates are a minimum.

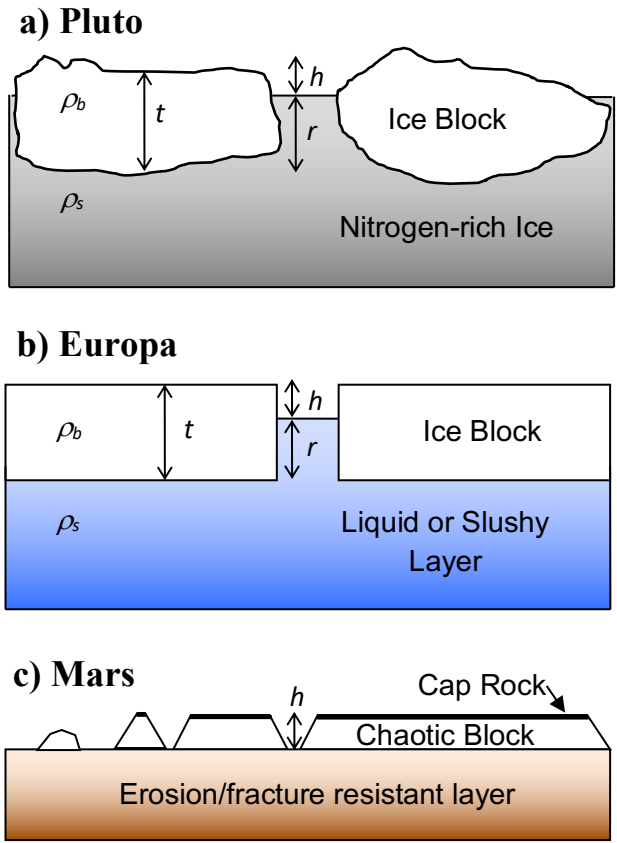

**Figure 7. Cartoons of chaotic mountain block profiles.** a) Parameters for estimation of block root depths (see Fig. 9) from isostasy are shown for Pluto. Not all blocks on Pluto are assumed to be floating (see text), but this still gives the setup for estimating how deep a root would be if the blocks were floating, or how deep SP would need to be to allow the blocks to float. A range of density contrasts is considered (see Fig. 9). b) Parameters for estimating ice shell thickness from isostasy are shown for Europa (see Fig. 8). Blocks on Europa may have reached an isostatic state shorty after formation if they were floating in a liquid or slushy layer before being frozen in place (see modelled ice and liquid compositions in Fig. 8). c) Possible lithospheric structure on Mars that could explain why chaotic block-top elevations are similar within a local region and why blocks seem to be limited in height. A resistant cap rock can protect the lower (and potentially less resistant) layers from weathering and erosion (see text). Note these cartoons are figurative and not to scale, nor do they represent all of the complexities of the lithospheres of each body.

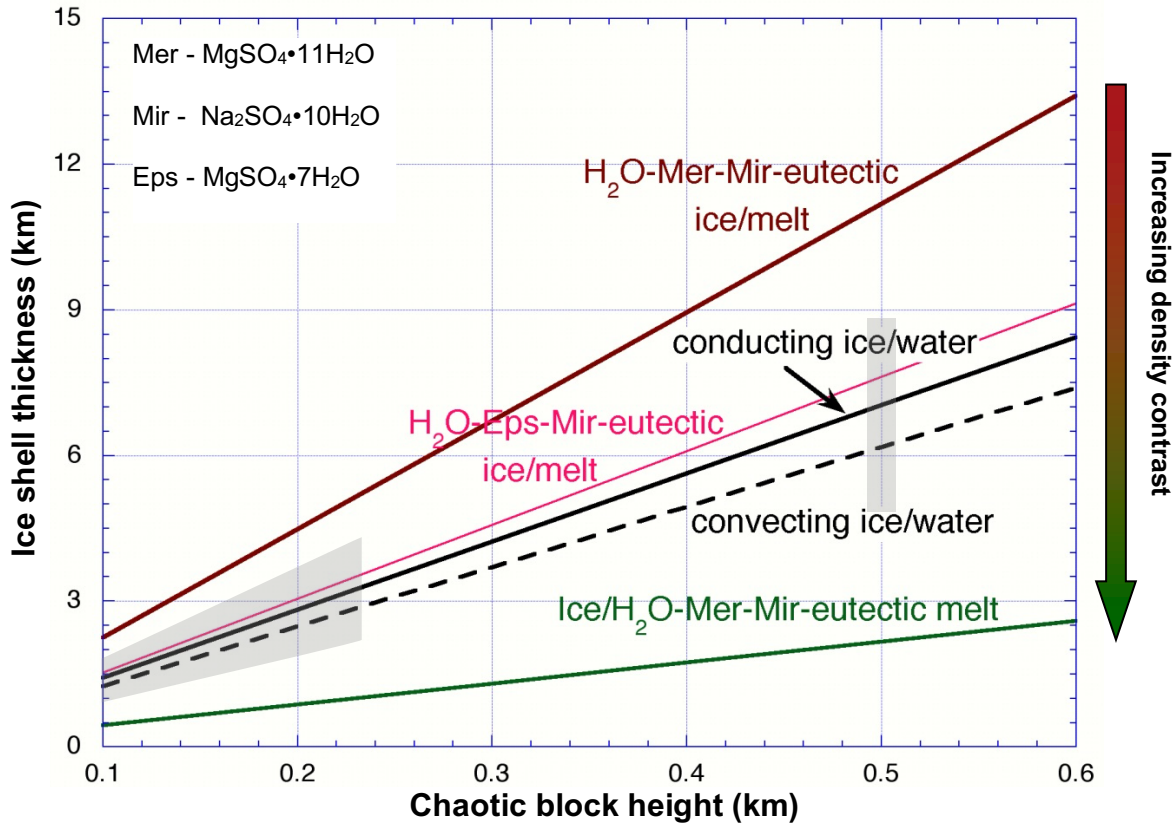

**Figure 8. Minimum ice shell thicknesses from isostasy for chaotic blocks on Europa.** Curves are plotted for a variety of possible ice shell/ocean compositions (chemistry as discussed in McKinnon and Zolensky, 2003) and show a range of solutions for more moderate to more extreme density contrasts. Using the more typical, untilted block heights of ~0.1 to 0.2 km, and less extreme compositions leads to ice shell thickness predictions of ~1 to 4 km (left-most grey box). The tallest blocks that appear to be untilted in the East RegMap region (~0.5 km in height) yields thicker minimum shell thickness estimates of ~5 to 9 km (right-most grey box). Figure adapted from Singer et al., 2021.

On Pluto, it is possible the chaotic blocks could have been partially or fully floating icebergs in the nitrogen ice sheet of SP, which could assist with destabilization or breakup/tilting (White et al., 2017; Howard et al., 2017). However, the distribution in **Figure 5a** does not match what is expected of floating blocks, where the blocks would be at similar heights (like the distribution of blocks in europan chaos; **Fig. 5b**) related to the density contrast between the blocks and the nitrogen-rich ice in SP. The increasing heights with increasing size imply that at least at the present moment most of the blocks are likely not floating. It is possible that the very largest blocks on may be reaching a maximum height of ~4 km (see **Figs. 3a, 5a**), but there are insufficient data points to infer if this could be indicative of layer thickness. Additionally, if the maximum height of blocks in mountain ranges in SP was determined to be around 4±1 km, the required root depth for the pure, solid water ice blocks to be floating in isostasy in the nitrogen rich ice sheet would be 40–50 km for the largest blocks. Unless the blocks are much less dense than water ice it is unlikely that the blocks are floating at a present time, as SP is estimated to be ~7–9 km deep near the center assuming impact origin (Nimmo et al., 2016; McKinnon et al., 2017) and is likely shallower near the edges.

On Pluto, the slopes of the block walls do not exceed ~34°, indicating they may also be at the angle of repose from erosive processes. The wall-slopes of Al-Idrisi Montes are generally shallower than those of the southern mountain groups, which is consistent with their more eroded appearance. However, if the erosion were similar to Mars (see text below), piles of finer material suggestive of erosional processes should be apparent at the base of the mountain blocks, but this is not the case. Although this type of observation may be limited by resolution, these piles might be seen in the higher resolution image strips for Pluto (~80–120 m px$^{-1}$). If finer material does erode off the block walls on Pluto, it may not stay at the base of mountains due to the movement of nitrogen ice (either through solid-state flow or sublimation and re-deposition). As demonstrated by Bertrand et al. (2018), high obliquity periods can induce intense polar summers associated with high sublimation rates in the northern part of the SP ice sheet, including near Al-Idrisi Montes. Their model suggests that over the last 2 million years, a removal of 1 km of ice due to sublimation occurred in this region. Balanced by glacial flow, this would have led to variation of elevation at the edges of the ice sheet up to 300 m during this period. This process would repeat every obliquity cycle and could explain the lack of erosional deposits at the base of the mountains.

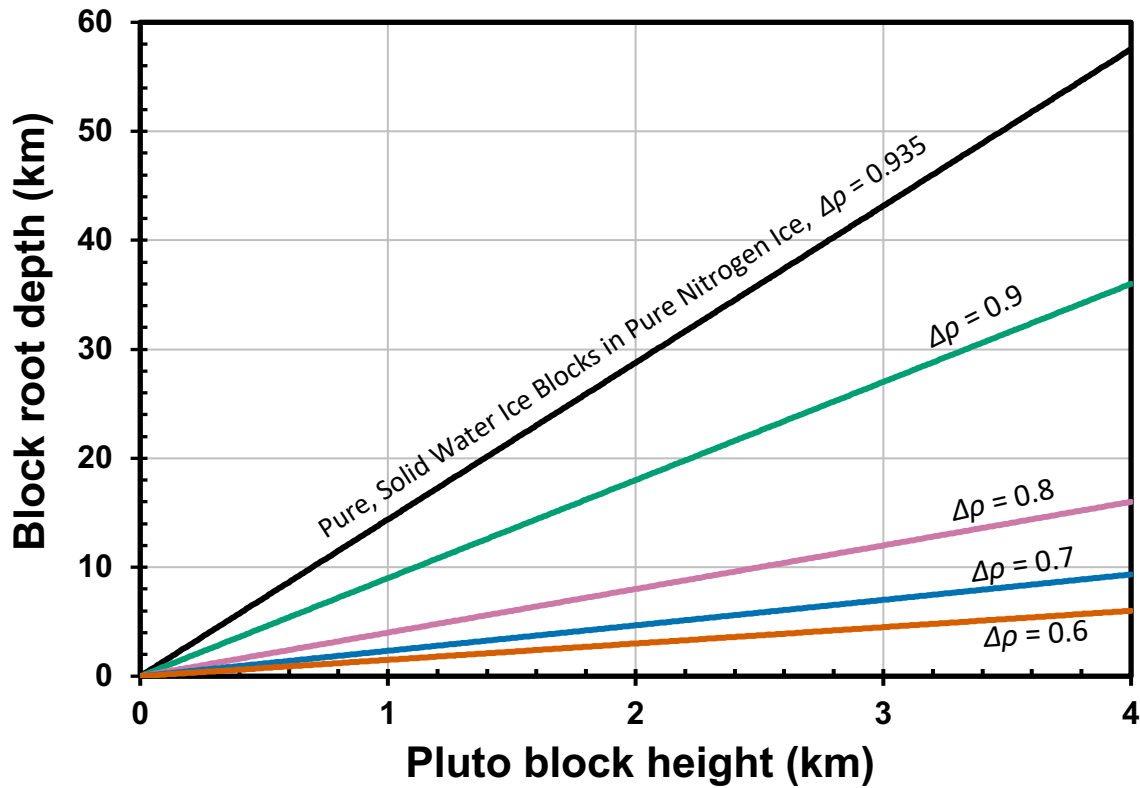

**Figure 9. Minimum root depth from isostasy for chaotic blocks on Pluto.** The minimum required root depth to support water ice blocks floating in the nitrogen-rich ice sheet of SP are plotted for a variety of possible density contrasts ($\Delta\rho$). Root depths at least 10s of km would be implied for the largest blocks except if the water ice is extensively fractured or porous.

For Mars a different process could lead to an estimate of layer thickness. Previous studies have used a combination of photogrammetry, spectroscopy, ground penetrating rater, gamma ray spectrometer, and limited surface-based reconnaissance to unveil the stratigraphic history on Mars (Thompson et al., 2016; Barker and Bhattacharya, 2018). The competence of lithologic layers could influence the maximum height of blocks, as different layers could be more resistant to erosional or deformational processes such as erosional downcutting, faulting and fracturing. The "surface tops" of martian blocks commonly matches the same elevation as the surrounding plateau (Meresse et al., 2008), and the maximum height of blocks in a region could be used to infer the relative layer thickness of a lithologic layer because erosional processes could have carved the weaker surface layers down to a more resistant layer. In certain regions blocks generally do not exceed above 1–1.5 km in height, compared to blocks in other regions reaching up to 2 km to around 2.5–2.8 km in height (**Fig. 5c**). Regional variation in the maximum height of blocks could be the result of subsidence (Meresse et al., 2008), spatial variability in the competence of lithologic layers, and/or differential exposure to erosional processes over time.

The slopes of chaotic block facture faces/walls on Mars generally appear eroded, and a pile-up of finer grains can be seen at the base of many blocks (even in 100 m px$^{-1}$ imaging). The slopes of the block walls on Mars are generally not greater than ~20–25 degrees (measured average slopes of chaos blocks and surrounding terrain boundaries are provided in **Table C2**, **Appendix C)**, and thus do not exceed the angle of repose for fine-grained, sandy material. The martian blocks have a characteristic diameter-height distribution where the left edge forms a sharp boundary (as seen in **Fig. 5c**). The blocks found along this boundary are more typically the crest-topped blocks, and the flat-topped blocks fill in the more distributed right-hand portions of the distribution (see example for Hydraotes Chaos in **Fig. C2**). Thus, the left edge of the martian plots gives an approximate proxy measurement of the tallest slopes found on the blocks walls. As an example, for Hydraotes Chaos (**Fig. C2**), the left edge of the distribution can be fit by a line with a slope ranging from 0.22 to 0.26 (depending on which blocks are chosen for the fit), similar to the measured slopes (**Table C2**). As can be seen in **Fig. 5c**, there is not much variation in the slope of the distribution edge for most regions. There are a few regions with a shallower slope (e.g., Aram and Iani Chaos) and a few regions where the left edge is not well expressed (e.g., Iamuna/Oxia, Ister, and Pyrrhae Chaos), but those regions have fewer or almost no crest-topped blocks, so the proxy does not work well in these regions.

## 6. CONCLUSIONS

We mapped blocks across chaotic terrain regions on Pluto, Europa and Mars, and quantified their morphology using measurements of diameter, height, and axial ratio. Mountain blocks in regions on Pluto and Mars display a positive linear relationship, whereas block heights on Europa do not scale with block size. Axial ratios provide information about the relative symmetry of blocks, which could be the result of patterns formed by fracture forming stress systems, and measurements of diameter and height could provide information about the lithologic structure of the crust on each body. Block heights on Pluto were used to calculate the required root depth for the blocks to be floating in isostasy in the nitrogen-rich SP basin. The tall heights of blocks on Pluto suggest that the larger blocks are not currently floating, as the estimated root depth of 10s of km would be too tall compared to the likely maximum depth of Sputnik basin along its perimeter of a few to 5 km. This may imply a thicker nitrogen ice cover in the past to have mobilized the mountain blocks in SP. On Europa block heights were used to estimate a *minimum* ice shell thickness of ~1 to 4 km at the time of the block formation, if the blocks reached an isostatic position after fracturing. On Mars we propose that block heights could be used to indicate a lithologic layer thickness if their apparent bases represent a more competent lithologic layer. Some regional variation in the maximum height of blocks on Mars (both in chaotic and fretted terrain) can be observed. Although the maximum height of blocks in fretted terrains is similar to that of chaotic terrain blocks, fretted terrain has a considerably wider diameter range than martian chaos. The larger blocks with lower heights in fretted terrain could be the result of erosional processes.

Data gathered by future spacecraft missions to Europa and Pluto, such as NASA's Europa Clipper mission, will provide much-needed additional constraints to enable further studies of these two bodies. Additional imaging data and topography of various geologic features, gravity data, and ice-penetrating radar will all yield additional insights into the lithospheric structures of Pluto and Europa. Meanwhile, additional mapping and modeling of the features we can observe with the current data will complement the results presented here and in previous works.

## ACKNOWLEDGEMENTS

New Horizons team members gratefully acknowledge funding from NASA's New Horizons project. We thank two anonymous reviewers for helpful comments and questions that improved this manuscript. We also thank Jeffrey Andrews-Hanna, Rebecca Thomas, and Alan Howard for informative discussions.

# APPENDIX A: Map views of Pluto, Europa, and Mars showing the locations of regions studied in the broader geologic context

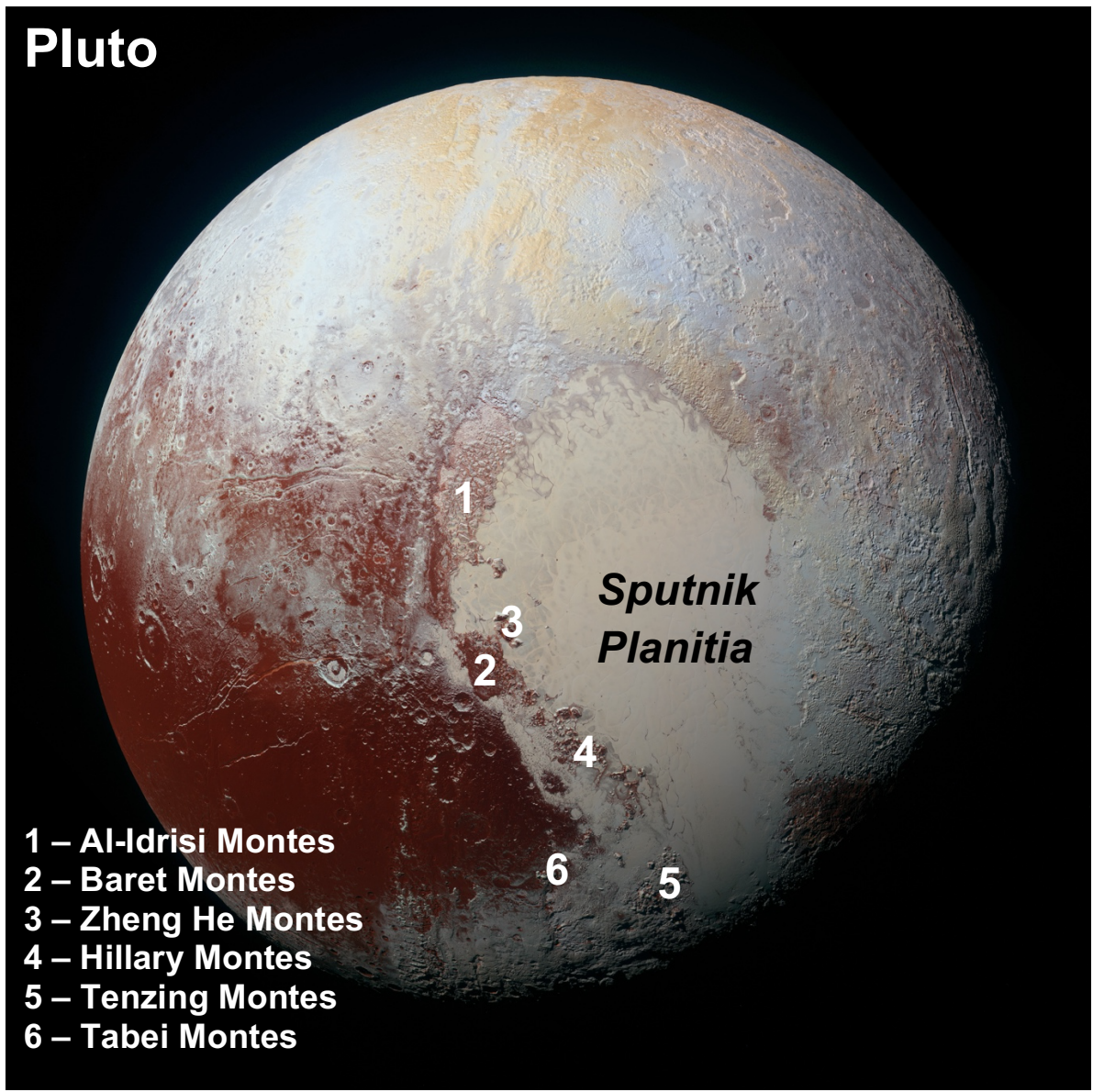

**Figure A1. Locations of montes studied on Pluto.** The chaotic mountain blocks on Pluto appear mainly on the western side of SP (Moore et al., 2016). Image courtesy: NASA/JHUAPL/SwRI.

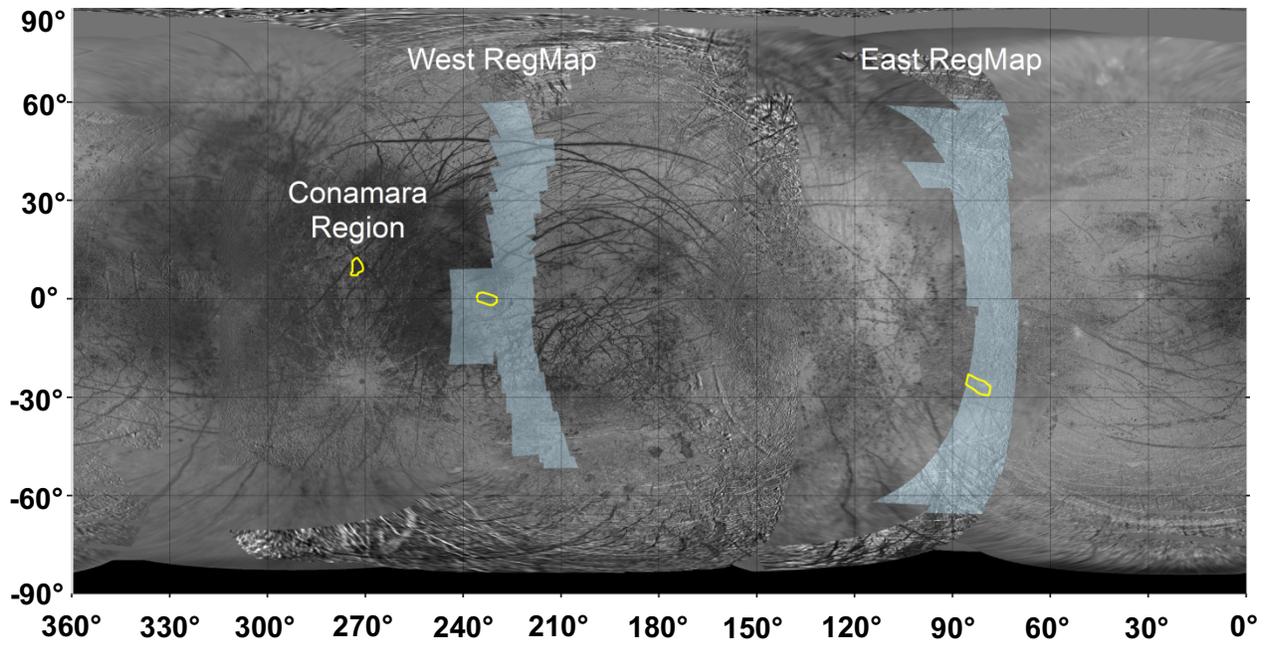

**Figure A2. Locations of mapped chaotic regions on Europa.** Light blue highlighting indicates east and west regional map coverage (~210 m px$^{-1}$). Yellow outlines indicate chaotic areas mapped for this study (see Fig. 1).

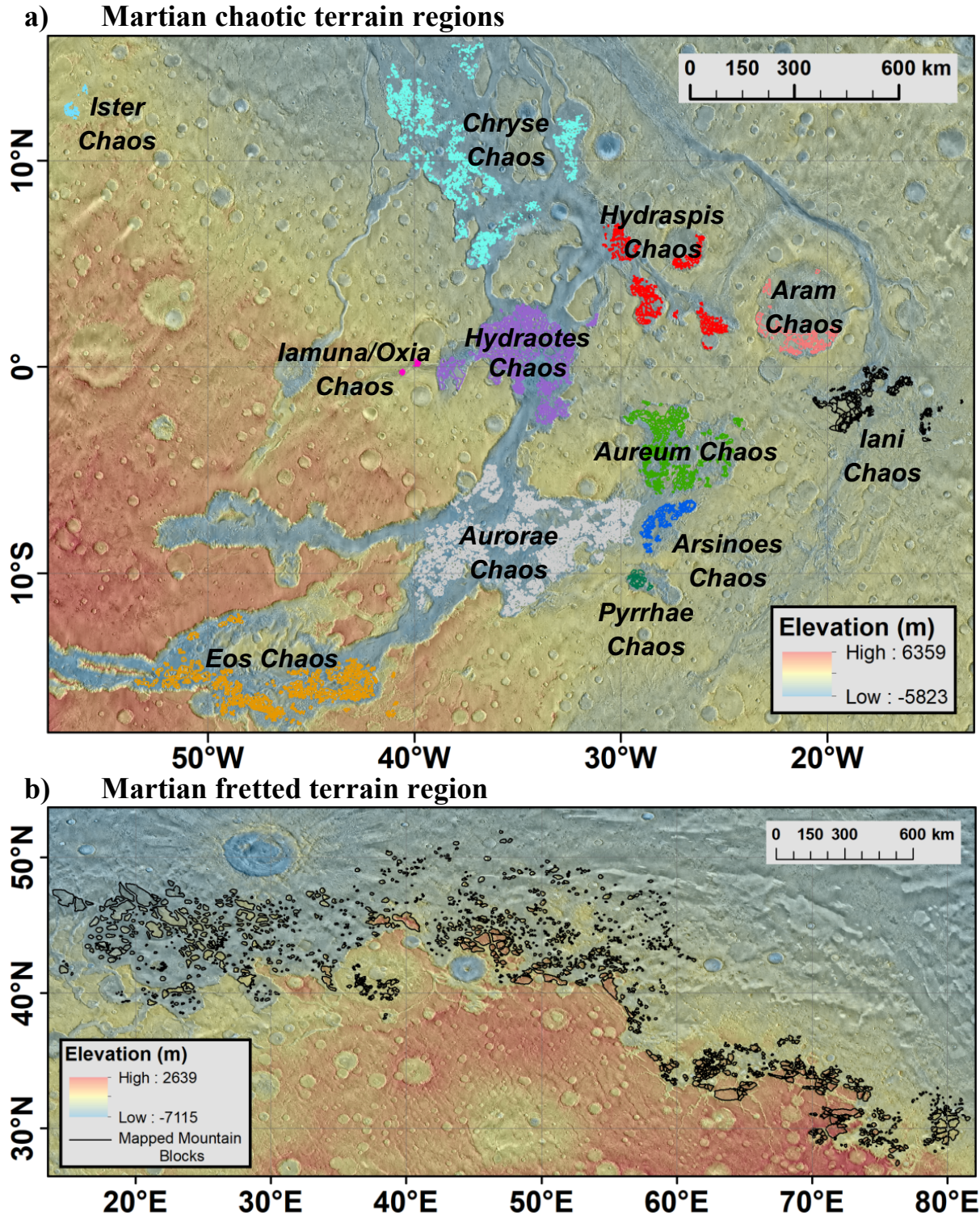

**Figure A3. Locations of mapped regions on Mars.** a) Locations of mapped chaotic terrain regions, and b) area of fretted terrain. Fretted terrain blocks were mapped along the lowland dichotomy boundary up to ~50° N. Elevation for martian chaotic and fretted terrain regions is displayed using MOLA topography (see Section 3.1 for discussion).

# APPENDIX B: Uncertainty in the estimated block height

As described in **Section 3.2**, the elevation of the surrounding terrain for each block was estimated from the mean elevation of the pixels that fall on the base outline of each block. The standard deviation of these pixel values was also computed and used to calculate the standard error of the mean elevation in that area ($\sigma\sqrt{N}$, where $N$ is the number of pixels). These standard errors were used to estimate the uncertainty on the measured height of the blocks through standard error propagation (e.g., Bevington and Robinson, 2003). In this case, we are using the maximum elevation value found withing a block polygon to calculate the height, and thus we keep that as an upper limit rather than a value with uncertainty (the vertical errors in the DEM values themselves are on the order of a few meters and thus do not play a strong role here; see **Section 3.1**). The uncertainty in the block heights is then the same as the uncertainty in the base elevation in this case. The resulting uncertainty, in meters, is given as a function of height for blocks on Pluto, Europa, and Mars (**Fig. B1**). The majority of the elevation points have uncertainties below 20 m on Pluto and Mars, and 10 m on Europa.

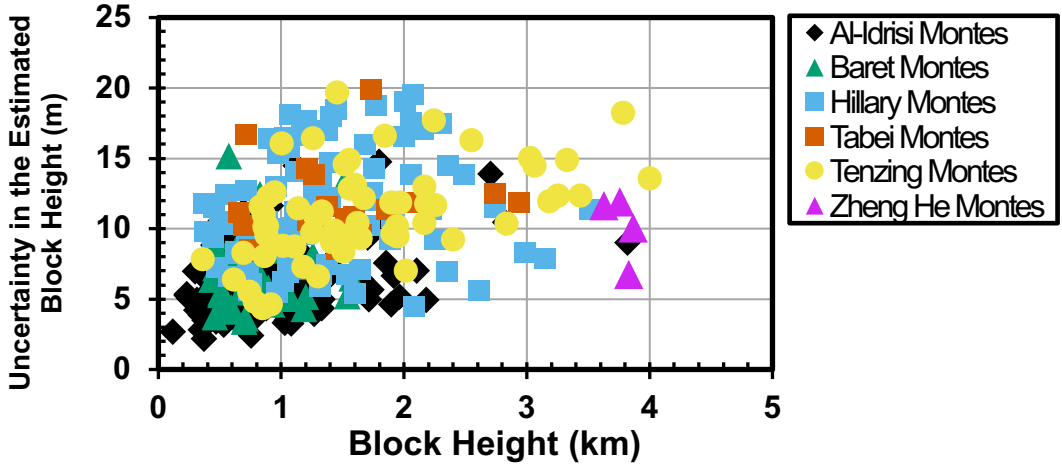
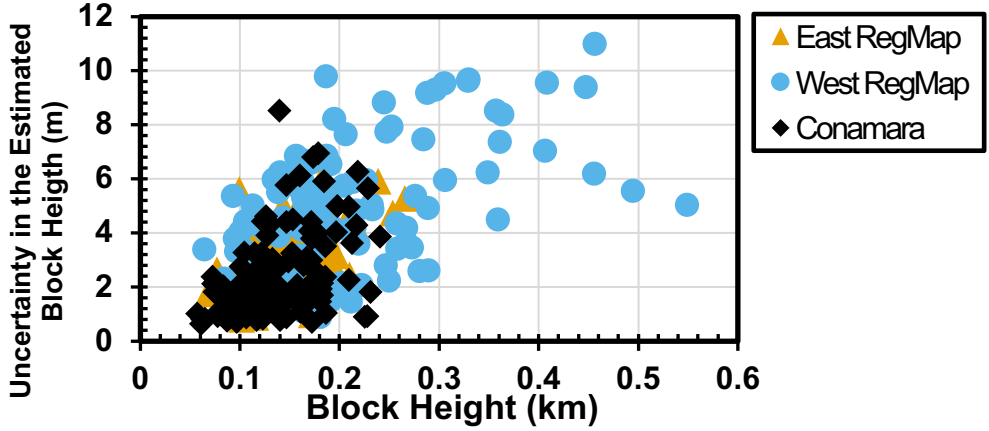
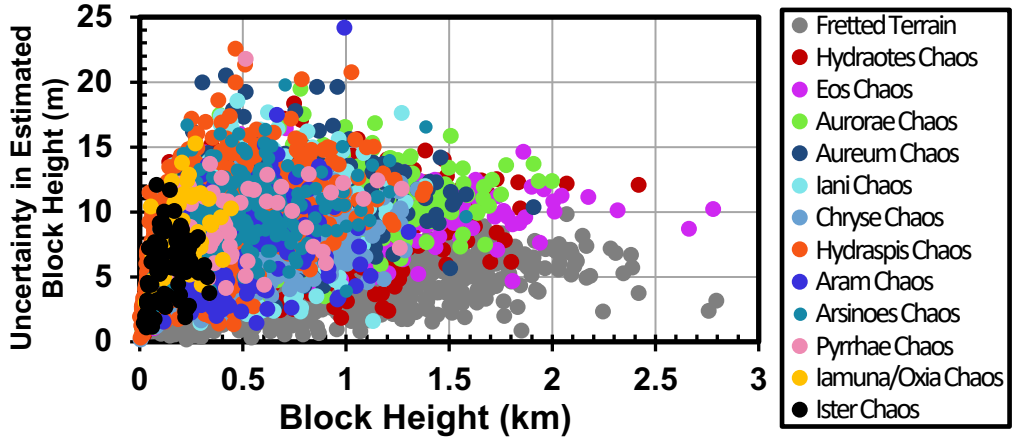

Figure B1. Uncertainty in the estimated height of a) plutonian, b) europan, and c) martian blocks.

# APPENDIX C: Block slope analysis for Pluto and Mars

To quantify the potential tilt of some mountain blocks on Pluto, we measured the average slope of the top of the block (dip angle) and the fracture face (see **Fig. C1a**). Although there are many blocks that appear to be tilted to varying degrees across our studied regions, we only measured blocks with tops that have surface textures resembling the surrounding upland terrain. Because of the uneven surfaces, we use an average slope estimate measured from the tallest point on the block to the lowest. **Table C1** provides information about each measured block and resulting slopes.

On Mars we measured the block slopes of chaotic terrain blocks and the surrounding terrain boundary (**Fig. C1b;** see **Table C2** for information about each slope measurement and results). Our slope calculations provide a rough average estimate of the side slope of each block, and were measured from the highest elevation point of each block to their apparent bottom where the block slopes flatten out and merge with the surrounding smooth valley floor.

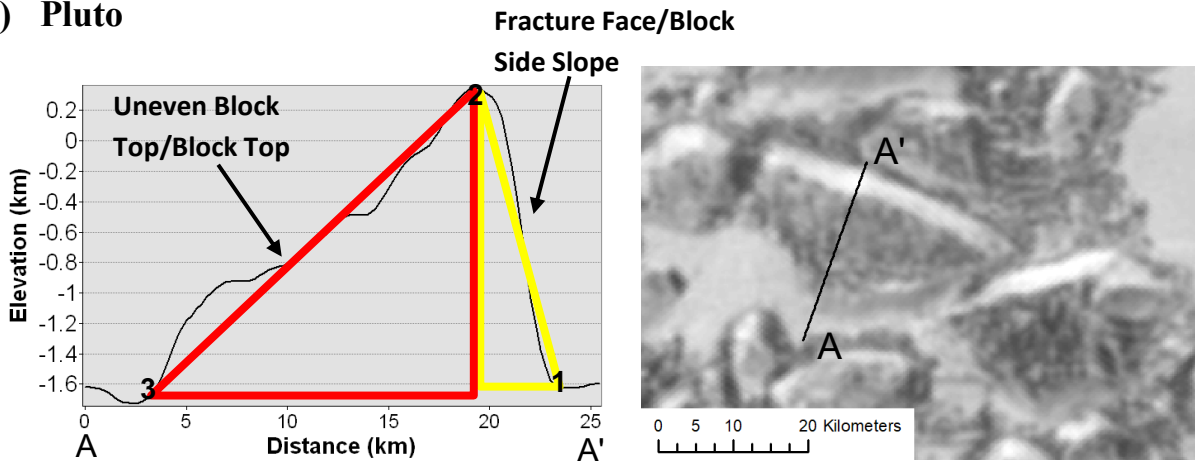

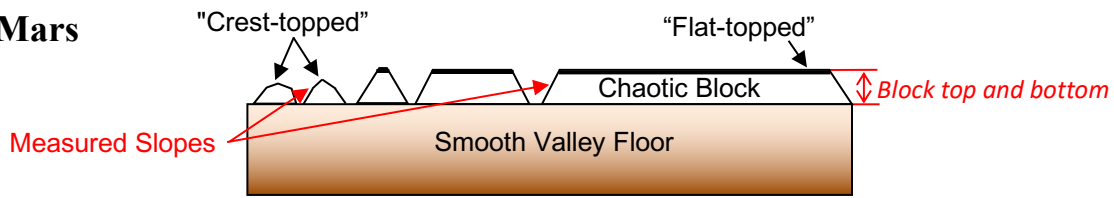

**Figure C1. Model figures for calculating the slope of the block top and its fracture face.** a) Our slope calculations provide a rough estimate of the dip of each block, with the assumption that the tops with surface textures resembling that of the adjacent surrounding uplands were previously horizontal before tilting during chaos formation or later modification. 1 and 3 represents the lowest point of each block, and 2 represents the tallest. Note extreme vertical exaggeration in left panel (thus the slope is not shown to scale). b) We measured the slope of the sides of martian chaotic terrain blocks for both flat-topped blocks, crest-topped blocks, and in some cases the slope of the nearby surrounding terrain boundary (e.g., boundary to the surrounding highland plateau) for comparison.

**Table C1. Average slopes of tilted blocks with exposed surface tops on Pluto.**

| Region | Block Location | ~Block Top Slope | ~Fracture Face/Block Side Slope |
|---|---|---|---|
| *Al-Idrisi Montes* | 33.9° N, 155.9° E | 3° | 17° |
| | 30° N, 154.5° E | 9° | 20° |
| | 29° N, 154° E | 6° | 27° |
| | 28.5° N, 156° E | 6° | 14° |
| | 26.9° N, 155° E | 8° | 13° |
| *Zheng He Montes* | 20.5° N, 159° E | 13° | 34° |
| | 18° N, 160° E | 7° | 30° |
| *Tenzing Montes* | 13.5° S, 175.5° E | 10° | 32° |
| | 14° S, 175° E | 4° | 28° |

**Table C2. Average slopes of chaos blocks and surrounding terrain boundaries on Mars.**

| Region | Slope Type | Location | ~Block Side Slope |
|---|---|---|---|
| *Chryse Chaos* | Surrounding terrain boundary | 9.1° N, 39.4° W | 19° |
| *Hydraotes Chaos* | Surrounding terrain boundary | 1.78° N, 36.8° W | 20° |
| | Surrounding terrain boundary | 2.6° N, 32.6° W | 17° |
| *Aurorae Chaos* | Surrounding terrain boundary | 4.9° S, 36.7° W | 23° |
| | Surrounding terrain boundary | 11.3° S. 37.6° W | 21° |
| *Eos Chaos* | Surrounding terrain boundary | 11.7° S, 50.9° W | 20° |
| | Surrounding terrain boundary | 13.7° S, 41.2° W | 24° |
| *Hydraotes Chaos* | Flat-topped block | 1° N, 34.6° W | 14° |
| | Flat-topped block | 1.1° N, 35.7° W | 25° |
| | Flat-topped block | 1.9° N, 36.0° W | 21° |
| | Flat-topped block | 1.9° N, 33.9° W | 12° |
| | Crest-topped block | 2.2° N, 34.6° W | 22° |
| | Crest-topped block | 0.1° S, 35.6° W | 20° |
| | Crest-topped block | 1.0° S, 33.3° W | 17° |
| *Aurorae Chaos* | Flat-topped block | 7.0° S, 31° W | 14° |
| | Flat-topped block | 7.0° S, 30° W | 19° |
| | Flat-topped block | 8.5° S, 32.4° W | 15° |
| | Crest-topped block | 6.8° S, 34.7° W | 15° |
| | Crest-topped block | 7.9° S, 32.3° W | 16° |
| | Crest-topped block | 9.0° S, 34° W | 17° |

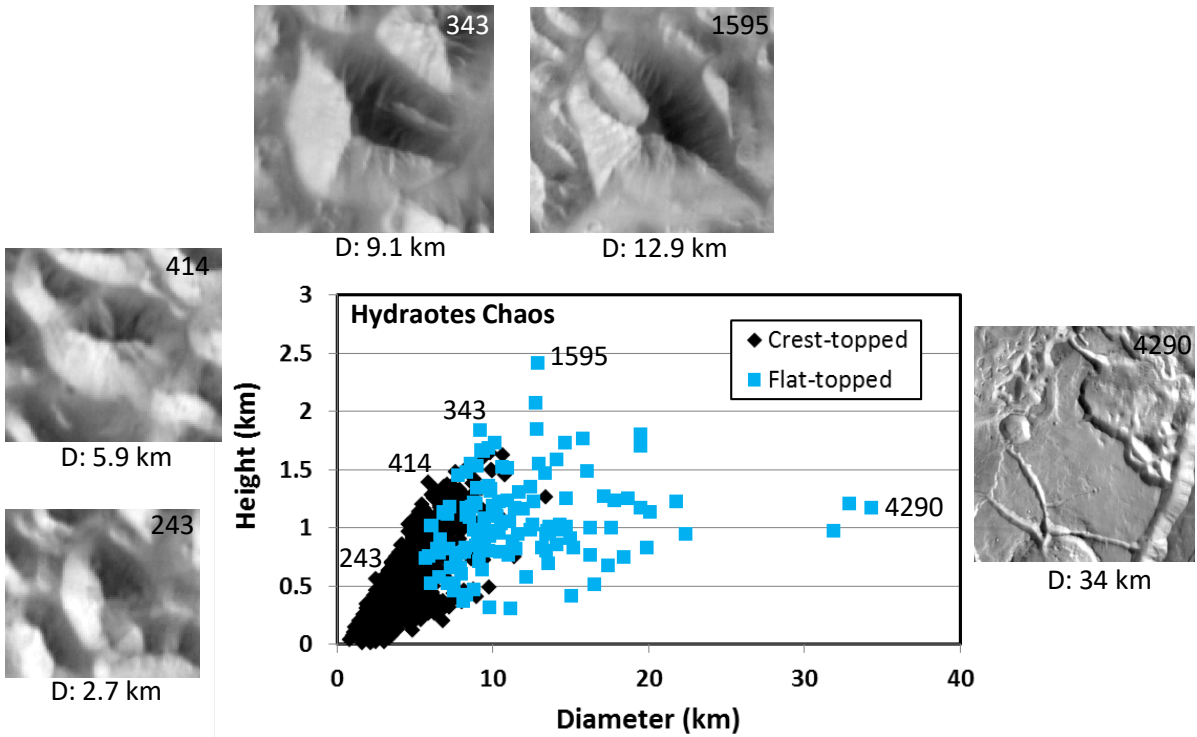

**Figure C2. Distribution of flat-topped and crest-topped blocks in Hydraotes Chaos, Mars.** We examined Hydraotes Chaos on Mars to confirm that the crest-topped blocks generally make up the sharp left edge of the diameter-height distribution seen in many of the martian chaos regions. As described in the text, this left edge is a proxy for the steepest slopes of the block walls on Mars. A few blocks are labeled with their feature IDs and are shown for reference with their equivalent diameters (D) noted.

# APPENDIX D: Fretted block height as a function of latitude

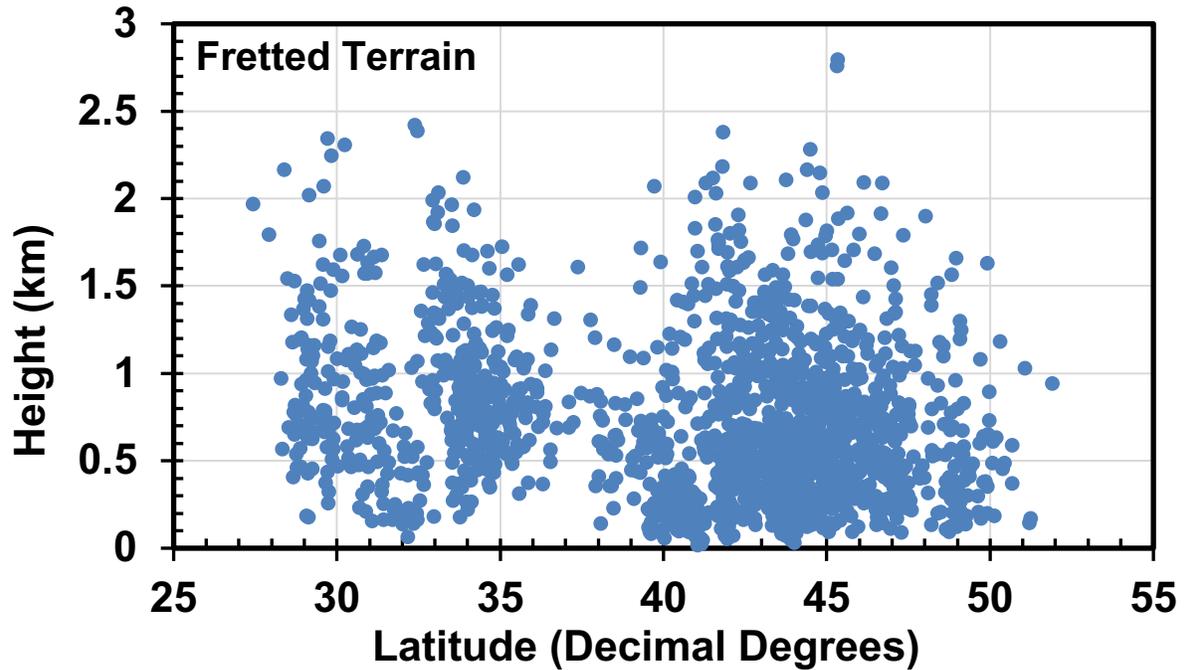

**Figure D1. Distribution of block height in fretted terrain with latitude.** We searched for any trends in the height of fretted blocks with increasing latitude, which is a proxy for distance from the dichotomy boundary. Although the overall elevation of all terrains decreases northward of the dichotomy boundary, as seen in Figure A3b, we do not find a strong trend in block height with latitude.

# APPENDIX E: Martian chaotic and fretted terrain regions plotted combined

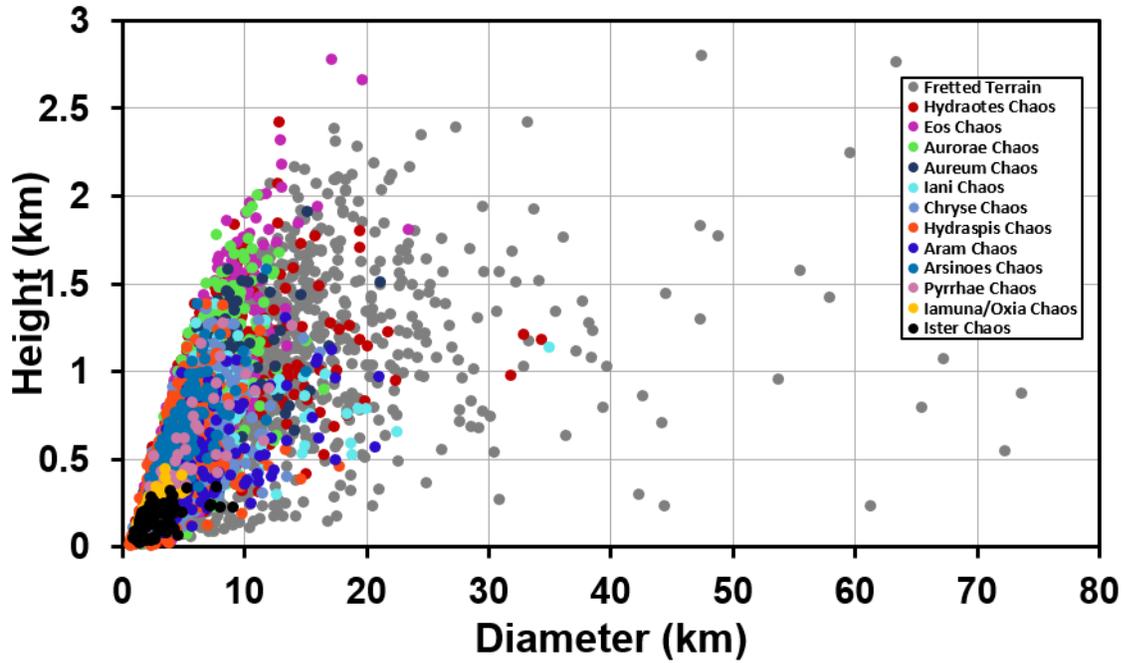

Figure E1. Comparison between feature topography with diameter for all regions studied across Mars.